\documentstyle[12pt]{article}

\makeatletter
\def\siml{\mathrel{\mathpalette\gl@align<}}
\def\simg{\mathrel{\mathpalette\gl@align>}}
\def\gl@align#1#2{\lower.6ex\vbox{\baselineskip\z@skip\lineskip\z@
 \ialign{$\m@th#1\hfill##\hfil$\crcr#2\crcr{\sim}\crcr}}}
\makeatother

\gdef\Feynmanlength{\setlength{\unitlength}{0.01pt}}  
\global\newcount\LINETYPE
\global\newcount\LINEDIRECTION
\global\newcount\LINECONFIGURATION
\newcommand{\LTYPE}{\LINETYPE}
\newcommand{\LDIR}{\LINEDIRECTION}
\newcommand{\LCONFIG}{\LINECONFIGURATION}
\global\LINETYPE=1  \global\LINEDIRECTION=0  
\global\LINECONFIGURATION=0
\global\newcount\fermion    \fermion=1
\global\newcount\scalar     \scalar=2
\global\newcount\photon     \photon=3
\global\newcount\gluon      \gluon=4
\global\newcount\SPECIAL    \SPECIAL=5
\gdef\N{0}  \gdef\NE{1}  \gdef\E{2}   \gdef\SE{3}
\gdef\S{4}  \gdef\SW{5}  \gdef\W{6}   \gdef\NW{7}
\global\newcount\REG            \global\REG=0
\global\newcount\FLIPPED        \global\FLIPPED=1
\global\newcount\CURLY          \global\CURLY=2
\global\newcount\FLIPPEDCURLY   \global\FLIPPEDCURLY=3
\global\newcount\FLAT           \global\FLAT=4
\global\newcount\FLIPPEDFLAT    \global\FLIPPEDFLAT=5
\global\newcount\CENTRAL        \global\CENTRAL=6
\global\newcount\FLIPPEDCENTRAL \global\FLIPPEDCENTRAL=7
             
\global\newcount\SQUASHEDGLUON  \global\SQUASHEDGLUON=8

\newcount\adjx \adjx=0
\newcount\adjy \adjy=0
\global\newdimen\BIGPHOTONS     \BIGPHOTONS=0pt  %  DEFAULT:  10 
\global\newdimen\THICKPHOTONS     \THICKPHOTONS=0pt  
\global\newdimen\THICKPHOTONSWITCH    \THICKPHOTONSWITCH=0pt
\gdef\THICKPHOTONTEST{
\THICKPHOTONSWITCH=0pt
\ifdim\THICKPHOTONS=0pt \relax
  \else \ifnum\LTYPE=3
           \ifnum\LDIR=2 \THICKPHOTONSWITCH=1pt \fi % THICK \E PHOTON
           \ifnum\LDIR=6 \THICKPHOTONSWITCH=1pt \fi % THICK \W PHOTON
        \fi
\fi
}  % end of THICKPHOTONTEST

\global\newcount\phantomswitch   \global\phantomswitch=0
\global\newcount\stemlength   \global\stemlength=275   
\global\newcount\absstemlength        % A copy of STEM length.
\global\newcount\stemlengthx          % FOR STEMS on particle lines
\global\newcount\stemlengthy          % FOR STEMS on particle lines
\newdimen\FRONTSTEM  \FRONTSTEM=0pt   % FOR STEMS
\newdimen\BACKSTEM   \BACKSTEM=0pt    % FOR STEMS
\newdimen\EITHERSTEM \EITHERSTEM=0pt  % FOR STEMS
\global\newcount\arrowlength                % FOR ARROWS
\global\newdimen\ATTIP   \global\ATTIP=0pt  % FOR ARROWS
\global\newdimen\ATBASE  \global\ATBASE=1pt % FOR ARROWS
\global\newcount\unitboxnumber  
\global\newcount\unitboxnumberpo  % One more than \unitboxnumber (in
\global\newcount\particlelengthx  
\gdef\plengthx{\particlelengthx}
\global\newcount\particlelengthy  
\gdef\plengthy{\particlelengthy}
\global\newcount\boxlengthx  
\global\newcount\boxlengthy  
\global\newcount\particleadjustx  
\global\newcount\particleadjusty  
\global\newcount\particlelength   % The LENGTH of a particle line BOX (x)
\global\newcount\particlefrontx
\gdef\pfrontx{\particlefrontx}
\global\newcount\PFRONTx
\global\newcount\particlefronty
\gdef\pfronty{\particlefronty}
\global\newcount\PFRONTy
\global\newcount\particlebackx
\gdef\pbackx{\particlebackx}
\global\newcount\particlebacky
\gdef\pbacky{\particlebacky}
\global\newcount\particlemidx
\gdef\pmidx{\particlemidx}
\global\newcount\particlemidy
\gdef\pmidy{\particlemidy}
\global\newcount\seglength  \global\newcount\gaplength
\global\gaplength=850  %default
\global\seglength=1416  % Length of each seg not including `ends' for
\global\newcount\Xone    \global\newcount\Yone    
\global\newcount\Xtwo    \global\newcount\Ytwo    
\global\newcount\Xthree  \global\newcount\Ythree  
\global\newcount\Xfour   \global\newcount\Yfour   % user co-ords
\global\newcount\Xfive   \global\newcount\Yfive   % user co-ords
\global\newcount\Xsix    \global\newcount\Ysix    
\global\newcount\Xseven  \global\newcount\Yseven  
\global\newcount\Xeight  \global\newcount\Yeight  
\newsavebox{\lastline}  %  Default name for an unnamed particle line.
\global\newcount\numlineparts   
\global\newcount\upperlineadjx  \upperlineadjx=0  %Default
\global\newcount\upperlineadjy  \upperlineadjy=0  %Default
\global\newcount\lowerlineadjx  \lowerlineadjx=0  %Default
\global\newcount\lowerlineadjy  \lowerlineadjy=0  %Default
\global\newcount\thirdlineadjx  \thirdlineadjx=0  %Default
\global\newcount\thirdlineadjy  \thirdlineadjy=0  %Default
\global\newcount\fourthlineadjx \fourthlineadjx=0  %Default
\global\newcount\fourthlineadjy \fourthlineadjy=0  %Default
\global\newcount\unitboxwidth   \unitboxwidth=1000%Default
\global\newcount\unitboxheight  \unitboxheight=0  %Default
\global\newcount\numupperunits  \numupperunits=8  %Default
\global\newcount\numlowerunits  \numlowerunits=8  %Default
\global\newcount\numthirdunits  \numthirdunits=8  %Default
\global\newcount\numfourthunits \numfourthunits=8  %Default
\global\newcount\fermioncount   \global\fermioncount=0
\global\newcount\scalarcount    \global\scalarcount=0
\global\newcount\photoncount    \global\photoncount=0
\global\newcount\gluoncount     \global\gluoncount=0
\global\newcount\SPECIALcount   \global\SPECIALcount=0
\global\newcount\vertexcount    \global\vertexcount=-1
\global\newcount\XDIR
\global\newcount\YDIR
\gdef\SETDIR{  % SETS THE DIRECTIONS
\ifcase\LDIR
     \global\XDIR=0  \global\YDIR=1   %\N  case.
\or  \global\XDIR=1  \global\YDIR=1   %\NE case.
\or  \global\XDIR=1  \global\YDIR=0   %\E  case.
\or  \global\XDIR=1  \global\YDIR=-1  %\SE case.
\or  \global\XDIR=0  \global\YDIR=-1  %\S  case.
\or  \global\XDIR=-1 \global\YDIR=-1  %\SW case.
\or  \global\XDIR=-1 \global\YDIR=0   %\W  case.
\or  \global\XDIR=-1 \global\YDIR=1   %\NW case.
\else\DIRECTERROR
\fi}  % END OF \SETDIR
\gdef\moduloeight#1{
\ifnum#1>7 \global\advance #1 by -8
\relax
\moduloeight#1
\relax
\else \relax
\fi}
\gdef\multroothalf#1{\global\multiply #1 by 7071 \global\divide #1 by 10000}
\gdef\negate#1{\global\multiply #1 by -1}

\gdef\slanttest(#1,#2){
\ifodd\LDIR
\multiply #1 by 7071  \divide #1 by 10000
\multiply #2 by 7071  \divide #2 by 10000
\fi
}
\gdef\gslanttest(#1,#2){
\ifodd\LDIR
\multroothalf#1
\multroothalf#2
\fi
}
\gdef\setplength{ % calcs length of particle line
\global\particlelengthx=\unitboxwidth
\global\particlelengthy=\unitboxheight
\global\multiply \particlelengthx by \unitboxnumber
\global\multiply \particlelengthy by \unitboxnumber
\global\advance \particlelengthx by \particleadjustx
\global\advance \particlelengthy by \particleadjusty
}
\gdef\boxlengthdefault{  % DEFAULT FOR BOX SIZES IN \drawas
\global\boxlengthx=\plengthx
\global\boxlengthy=\plengthy
\ifnum\plengthx<0 \global\multiply\boxlengthx by -1 \fi
\ifnum\plengthy<0 \global\multiply\boxlengthy by -1 \fi
}
\gdef\rearcoords{  
\global\particlebacky=\particlefronty
\global\particlebackx=\particlefrontx
\global\advance \particlebackx by \particlelengthx
\global\advance \particlebacky by \particlelengthy
}
\gdef\midcoords{  
\global\particlemidy=\particlefronty
\global\particlemidx=\particlefrontx
\global\stemlengthx=\particlelengthx  
\global\stemlengthy=\particlelengthy
\global\divide\stemlengthx by 2
\global\divide\stemlengthy by 2
\global\advance \particlemidx by \stemlengthx
\global\advance \particlemidy by \stemlengthy
}
\gdef\setparticle{\setplength\rearcoords\midcoords\boxlengthdefault}  
\gdef\setcoords(#1,#2,#3)(#4,#5,#6)[#7,#8]{
\global\upperlineadjx=#1
\global\lowerlineadjx=#2
\global\thirdlineadjx=#3
\global\upperlineadjy=#4
\global\lowerlineadjy=#5
\global\thirdlineadjy=#6
\global\unitboxwidth=#7
\global\unitboxheight=#8
}
\gdef\drawoldpic#1(#2,#3){  % DRAWS PRE-SAVED PICTURE
\global\particlefrontx=#2
\global\particlefronty=#3
\rearcoords
\midcoords
\put(#2,#3){\usebox{#1}}
}
\gdef\drawsavedline`#1' as #2[#3#4](#5,#6)[#7]{
\global\LINETYPE=#2
\global\LINEDIRECTION=#3
\global\LINECONFIGURATION=#4
\global\particlefrontx=#5
\global\particlefronty=#6
\global\unitboxnumber=#7
\selectcase
\rearcoords% moved from before selectcase.
\midcoords
\ifnum\phantomswitch=0 \drawas{#1}\fi
}

\gdef\drawas#1{
\global\savebox{#1}(\boxlengthx,\boxlengthy){
\setlength{\unitlength}{0.01pt}
\begin{picture}(\boxlengthx,\boxlengthy)
\multiput(\upperlineadjx,\upperlineadjy)(\unitboxwidth,\unitboxheight)
{\numupperunits}{\upperunitbox}
\ifnum\numlineparts > 1  %  If the line needs 2 parts per unit or more
\multiput(\lowerlineadjx,\lowerlineadjy)(\unitboxwidth,\unitboxheight)
{\numlowerunits}{\lowerunitbox}
\fi
\ifnum\numlineparts > 2  %  If the line needs 3 parts per unit or more
\multiput(\thirdlineadjx,\thirdlineadjy)(\unitboxwidth,\unitboxheight)
{\numthirdunits}{\thirdunitbox}
\fi
\ifnum\numlineparts > 3  %  If the line needs 4 parts per unit or more
\multiput(\fourthlineadjx,\fourthlineadjy)(\unitboxwidth,\unitboxheight)
{\numfourthunits}{\lowerunitbox}
\fi
\end{picture} }
\global\PFRONTx=\pfrontx  \global\PFRONTy=\pfronty   %save this value
\SETFRONTSTEM
\THICKPHOTONTEST
\ifdim\THICKPHOTONSWITCH=1pt\global\advance\PFRONTy by 20  \fi
\put(\PFRONTx,\PFRONTy) {\usebox{#1}}   %\pfrontX,Y=\particlefrontx,y
\ifdim\THICKPHOTONSWITCH=1pt
\global\advance\PFRONTy by -40
\put(\PFRONTx,\PFRONTy) {\usebox{#1}}   
\global\advance \PFRONTy by 20  
\fi  %End of \ifdim\THICKPHOTONSWITCH=1
\SETBACKSTEM
\seglength=1416   \gaplength=850   % Re-set \SCALR defaults.
}
\gdef\drawandsaveline`#1' as #2[#3#4](#5,#6)[#7]{
\global\newsavebox{#1}
\drawsavedline`#1' as #2[#3#4](#5,#6)[#7]
}

\gdef\drawline#1[#2#3](#4,#5)[#6]{   % Draw line but don't name it.
\drawsavedline`\lastline' as #1[#2#3](#4,#5)[#6]}

\gdef\TYPEERROR{\message{*** ERROR IN PARTICLE TYPE 
SELECTION ***}
\message{+++ Try with line type \fermion,\scalar,\photon,\gluon
(see manual) +++}\SETERR}
\gdef\DIRECTERROR{\SETERR\message{*** ERROR IN PARTICLE 
DIRECTION SELECTION
***}
\message{+++ Try again with direction N, NE, E, SE  etc. or see 
manual +++}}
\gdef\UNIMPERROR{\message{*** ERROR IN PARTICLE OPTIONS 
SELECTION ***}
\message{
+++ The requested options combination has not yet been implemented 
+++}\SETERR}
\gdef\SETERR{\gdef\upperunitbox{{\tiny Error}}  
\gdef\lowerunitbox{\relax}
\gdef\thirdunitbox{\relax}
}
\gdef\neglengthcheck{\ifnum\unitboxnumber < 1
\message{   *** ERROR:  PARTICLE OF NEGATIVE OR ZERO 
LENGTH REQUESTED. ***   }
\message{   ***         TAKING ABSOLUTE VALUE. ***   }\negate
\unitboxnumber\fi}
\gdef\selectcase{
\neglengthcheck   %  check for particles of negative length.
\SETDIR
\ifcase\LINETYPE
\TYPEERROR  % \LINETYPE=0 case.
\or \selectfermion  % \LINETYPE=1 case.
\or \selectscalar   % \LINETYPE=2 case.
\or \selectphoton   % \LINETYPE=3 case.
\or \selectgluon    % \LINETYPE=4 case.
\or \selectspecial  % \LINETYPE=5 case.
\else \TYPEERROR \fi  }
\gdef\selectfermion{
\ifnum\fermioncount=0

\global\newcount\fermionlength  %  THE TOTAL FERMION LINE LENGTH.
\global\newcount\fermionlengthx
\global\newcount\fermionlengthy
\global\newcount\fermionfrontx  %}(x,y) co-ord of left of fermion
\global\newcount\fermionfronty  %}
\global\newcount\fermionbackx
\global\newcount\fermionbacky
\gdef\ALLfermion{  % READ IN FROM FEYNMAN \selectfermion
\global\fermionfrontx=\particlefrontx \global\fermionfronty=\particlefronty
\ifnum\unitboxnumber > 50000
\message{   *** WARNING *** Fermion of length
\the\unitboxnumber\space requested ***   }
\ifnum\unitboxnumber > 80000
\message{   *** Reducing fermion length to 30000 (max 80000) ***   }
\global\unitboxnumber=30000 \fi \fi  % end of length error
\global\fermionlength=\unitboxnumber % The TOTAL line length
\global\particleadjustx=0   \global\particleadjusty=0 %Default
\global\numlineparts = 1    \global\numupperunits=1
\global\upperlineadjx=-200  \global\upperlineadjy=0
\global\fermionlengthx=\fermionlength    
\global\fermionlengthy=\fermionlength
\gslanttest(\fermionlengthx,\fermionlengthy)  
\global\multiply\fermionlengthx by \XDIR  %  In keeping with photons and
\global\multiply\fermionlengthy by \YDIR  %  In keeping with photons and
\global\unitboxheight=\fermionlengthy   
\global\unitboxwidth=\fermionlengthx
\global\advance \fermionlengthx by \particleadjustx
\global\advance \fermionlengthy by \particleadjusty
\global\particlelengthx=\fermionlengthx
\global\particlelengthy=\fermionlengthy
\boxlengthdefault    \rearcoords    \midcoords
\global\fermionbackx=\particlebackx     
\global\fermionbacky=\particlebacky
\ifcase\LINECONFIGURATION  %\REG case
\ifnum\XDIR=0
\gdef\upperunitbox{\line(\XDIR,\YDIR){\boxlengthy}} %\N or \S
\else
\gdef\upperunitbox{\line(\XDIR,\YDIR){\boxlengthx}}
\fi
\else \UNIMPERROR
\fi
}

\fi
\global\advance\fermioncount by 1  % Counts number of fermions drawn.
\ALLfermion
}
\gdef\selectscalar{
\ifnum\scalarcount=0
\newcount\scalarlength
\newcount\scalarlengthx
\newcount\scalarlengthy
\newcount\scalarfrontx  %}(x,y) co-ord of left of scalar
\newcount\scalarfronty  %}
\newcount\scalarbackx
\newcount\scalarbacky
\gdef\ALLscalar{
\global\scalarfrontx=\particlefrontx   
\global\scalarfronty=\particlefronty   
\numlineparts = 1      \numupperunits=\unitboxnumber
\ifcase\LINECONFIGURATION
\global\upperlineadjx=-200     \global\upperlineadjy=0
\slanttest(\seglength,\gaplength)   %SEE FEYNMAN22.TEX.
\gdef\upperunitbox{\line(\XDIR,\YDIR){\seglength}}
\else \UNIMPERROR % etc.
\fi
\global\unitboxwidth=\seglength  \global\advance\unitboxwidth by 
\gaplength
\global\multiply \unitboxwidth by \XDIR
\global\unitboxheight=\seglength  \global\advance\unitboxheight 
by \gaplength
\global\multiply \unitboxheight by \YDIR
\global\particleadjustx=\gaplength \global\multiply\particleadjustx by \XDIR
\global\particleadjusty=\gaplength \global\multiply\particleadjusty by \YDIR
\negate\particleadjustx   \negate\particleadjusty   
\setparticle  %SCALAR8
\global\scalarlengthx=\particlelengthx  %SCALAR8
\global\scalarlengthy=\particlelengthy  %SCALAR8
\ifnum\boxlengthx > 50000
\message{   *** WARNING *** Scalar of length in excess of 50000cp
requested!}\fi
\ifnum\boxlengthy > 50000
\message{   *** WARNING *** Scalar of length in excess of 50000cp
requested!}\fi
\global\scalarbackx=\pbackx      \global\scalarbacky=\pbacky   %SCALAR8
}

*****************************************************************************

\fi
\global\advance\scalarcount by 1  % Counts number of scalars drawn.
\ALLscalar
}
\gdef\selectphoton{   % RECURSIVELY RE-DEFINED IN 
\ifnum\photoncount=0

\newcount\numwiggles    \newcount\numwigglespo
\global\newcount\photonlengthx
\global\newcount\photonlengthy
\global\newcount\photonfrontx  %}(x,y) co-ord of left of photon
\global\newcount\photonfronty  %}
\global\newcount\photonbackx
\global\newcount\photonbacky
\newcount\halfwigglelength
\global\font\Twelverom=cmr12
\global\font\Tenrom=cmr10
\gdef\Lbr{{\Twelverom(}}   \gdef\Rbr{{\Twelverom)}}
\gdef\SLbr{{\Tenrom(}}     \gdef\SRbr{{\Tenrom)}}
\gdef\Smile{{\large$\smile$}}  % Default for 10 and 11-point documents.
\gdef\Frown{{\large$\frown$}}  % Default for 10 and 11-point documents.
\ifdim\BIGPHOTONS>0pt  \gdef\Smile{$\smile$} \gdef\Frown{$\frown$} \fi
\gdef\selectphoton{   % RECURSIVELY RE-DEFINED HERE.  
\global\advance\photoncount by 1  % Counts number of photons drawn.
\global\photonfrontx=\particlefrontx   
\global\photonfronty=\particlefronty   
\ifnum\unitboxnumber > 50
\message{   *** WARNING *** Photon with
\the\unitboxnumber\space half-wiggles requested ***   }
\ifnum\unitboxnumber > 150
\message{   *** Reducing photon length to 10 half-wiggles (max 150) ***   }
\ifnum\unitboxnumber > 1000
\message{   *** Probable Cause:  Photon selected instead of Fermion ***   }
\fi \global\unitboxnumber=10 \fi \fi  % end of length error
\numwiggles=\unitboxnumber
\divide\numwiggles by 2
\global\unitboxnumberpo=\numwiggles 
\global\multiply \unitboxnumberpo by -1
\numwigglespo=\unitboxnumber
\advance\numwigglespo by \unitboxnumberpo 
\global\numlineparts = 2  % DEFAULT                
\global\numupperunits=\numwigglespo  % DEFAULT
\global\numlowerunits=\numwiggles  % DEFAULT
\particleadjustx=0  %DEFAULT
\particleadjusty=0  %DEFAULT
\ifcase\LINEDIRECTION
     \Nphoton    %\LINEDIRECTION=0 (NORTH) CASE
\or  \NEphoton   % 1 case
\or  \Ephoton    % 2 case...horizontal photon.
\or  \SEphoton   % .
\or  \Sphoton    % .
\or  \SWphoton   % .
\or  \Wphoton    % .
\or  \NWphoton   % 7 case
\else\DIRECTERROR \fi
\setplength
\global\divide\plengthx by 2  \global\divide\plengthy by 2
\rearcoords  \boxlengthdefault   \midcoords
\global\photonbackx=\pbackx  %PHOTONSETUP26
\global\photonbacky=\pbacky  %PHOTONSETUP26
\global\photonlengthx=\plengthx  %PHOTONSETUP26
\global\photonlengthy=\plengthy  %PHOTONSETUP26
}
\gdef\SETUNITBOX(##1)[##2][##3]{ % For slanted photons only.
\gdef\upperunitbox{\oval(##1,##1)[##2]}
\gdef\lowerunitbox{\oval(##1,##1)[##3]}
}
\gdef\Nphoton{  % VERTICAL PHOTONS
\ifcase\LINECONFIGURATION  %\REG case
\setcoords(-490,-250,0)(260,1250,0)[0,2000]
\gdef\upperunitbox{\SLbr}   \gdef\lowerunitbox{\SRbr}
\particleadjusty=10
\or % \FLIPPED case
\setcoords(-271,-501,0)(250,1250,0)[0,2000]
\gdef\upperunitbox{\SRbr}   \gdef\lowerunitbox{\SLbr}
\or %\CURLY case (a bit shorter).
\particleadjusty=0
\setcoords(-501,-351,0)(300,1400,0)[0,2200]
\gdef\upperunitbox{\Lbr}   \gdef\lowerunitbox{\Rbr}
\or %\FLIPPEDCURLY case.
\setcoords(-353,-499,0)(300,1400,0)[0,2200]
\gdef\upperunitbox{\Rbr}   \gdef\lowerunitbox{\Lbr}
\or % \FLAT case.  Flatter and shorter than \CURLY.
\setcoords(-481,-371,0)(280,1300,0)[0,2000]
\gdef\upperunitbox{\Lbr}   \gdef\lowerunitbox{\Rbr}
\particleadjusty=150
\ifnum\numwiggles=\number\numwigglespo \particleadjustx=-50 \fi
\or %\FLIPPEDFLAT case.  \LINECONFIGURATION=5.
\setcoords(-321,-391,0)(280,1300,0)[0,2000]
\gdef\upperunitbox{\Rbr}   \gdef\lowerunitbox{\Lbr}
\particleadjusty=150
\ifnum\numwiggles=\number\numwigglespo \particleadjustx=80 \fi
\or % \LONGPHOTON
\setcoords(-490,-260,0)(300,1500,0)[0,2400]
\gdef\upperunitbox{\Lbr}   \gdef\lowerunitbox{\Rbr}
\or % \FLIPPEDLONGPHOTON
\setcoords(-301,-531,0)(300,1500,0)[0,2400]
\gdef\upperunitbox{\Rbr}   \gdef\lowerunitbox{\Lbr}
\else \UNIMPERROR
\fi
}
\gdef\NEphoton{    
\ifcase\LINECONFIGURATION  %\REG case
\setcoords(425,425,0)(1250,0,0)[1250,1250]       \SETUNITBOX(1250)[br][tl]
\ifnum\numwigglespo > \number \numwiggles \particleadjustx=15 \fi
\or % \FLIPPED case
\setcoords(1050,-200,0)(625,625,0)[1250,1250]    \SETUNITBOX(1250)[tl][br]
\ifnum\numwigglespo > \number \numwiggles \particleadjustx=25 \fi
\or % \CURLY case.
\setcoords(500,500,0)(1400,0,0)[1400,1400]       \SETUNITBOX(1400)[br][tl]
\or % \FLIPPEDCURLY case
\setcoords(1200,-200,0)(700,700,0)[1400,1400]    \SETUNITBOX(1400)[tl][br]
\or % \FLAT case
\setcoords(400,400,0)(1200,0,0)[1200,1200]       \SETUNITBOX(1200)[br][tl]
\or % \FLIPPEDFLAT case
\setcoords(1000,-200,0)(600,600,0)[1200,1200]    \SETUNITBOX(1200)[tl][br]
\else \UNIMPERROR
\fi
\numupperunits=\numwiggles   \numlowerunits=\numwigglespo
}
\gdef\Ephoton{    %  EASTWARD  HORIZONTAL PHOTONS
\ifcase\LINECONFIGURATION  % REG case
\setcoords(-285,715,0)(-150,-400,0)[2005,0]
\gdef\upperunitbox{\Frown}   \gdef\lowerunitbox{\Smile}
\or  % \FLIPPED case
\setcoords(-285,715,0)(-420,-170,0)[2005,0]
\gdef\upperunitbox{\Smile}   \gdef\lowerunitbox{\Frown}
\else \UNIMPERROR
\fi
\particleadjustx=-15 % Lengths are in centipoints.
}
\gdef\SEphoton{   
\ifcase\LINECONFIGURATION  %\REG case
\setcoords(-200,1050,0)(-625,-625,0)[1250,-1250] 
\SETUNITBOX(1250)[tr][bl]
\ifnum\numwigglespo > \number \numwiggles \particleadjustx=25 \fi
\or % \FLIPPED case
\setcoords(425,425,0)(0,-1250,0)[1250,-1250]     
\SETUNITBOX(1250)[bl][tr]
\ifnum\numwigglespo > \number \numwiggles \particleadjustx=15 \fi
\or % \CURLY case.
\setcoords(-200,1200,0)(-700,-700,0)[1400,-1400] 
\SETUNITBOX(1400)[tr][bl]
\or % \FLIPPEDCURLY case
\setcoords(500,500,0)(0,-1400,0)[1400,-1400]     
\SETUNITBOX(1400)[bl][tr]
\or % \FLAT case
\setcoords(-200,1000,0)(-600,-600,0)[1200,-1200] 
\SETUNITBOX(1200)[tr][bl]
\particleadjustx=-20
\or % \FLIPPEDFLAT case
\setcoords(420,420,0)(0,-1200,0)[1200,-1200]     
\SETUNITBOX(1200)[bl][tr]
\particleadjustx=40
\else \UNIMPERROR
\fi
}
\gdef\Sphoton{  % DOWN, DOWN VERTICAL PHOTONS
\ifcase\LINECONFIGURATION  %\REG case
\setcoords(-252,-490,0)(-740,-1740,0)[0,-2000]
\gdef\upperunitbox{\SRbr}   \gdef\lowerunitbox{\SLbr}
\or % \FLIPPED case
\setcoords(-490,-260,0)(-740,-1740,0)[0,-2002]
\gdef\upperunitbox{\SLbr}   \gdef\lowerunitbox{\SRbr}
\or %\CURLY case (a bit shorter).
\setcoords(-299,-449,0)(-870,-1970,0)[0,-2200]
\gdef\upperunitbox{\Rbr}    \gdef\lowerunitbox{\Lbr}
\particleadjusty=-95
\or %\FLIPPEDCURLY case.
\setcoords(-517,-371,0)(-900,-2000,0)[0,-2200]
\gdef\upperunitbox{\Lbr}    \gdef\lowerunitbox{\Rbr}
\particleadjusty=-165
\or % \FLAT case.  Flatter and shorter than \CURLY.  
\setcoords(-299,-409,0)(-885,-1905,0)[0,-2000]
\gdef\upperunitbox{\Rbr}   \gdef\lowerunitbox{\Lbr}
\particleadjustx=50     \particleadjusty=-380
\ifodd\unitboxnumber\relax\else\particleadjustx=250 
\particleadjusty=-400 \fi
\or %\FLIPPEDFLAT case.  \LINECONFIGURATION=5.
\setcoords(-519,-449,0)(-900,-1920,0)[0,-2000]
\gdef\upperunitbox{\Lbr}   \gdef\lowerunitbox{\Rbr}
\particleadjusty=-370
\ifodd\unitboxnumber\relax\else\particleadjustx=-240 
\particleadjusty=-400 \fi
\or % \LONGPHOTON
\gdef\upperunitbox{\Rbr}   \gdef\lowerunitbox{\Lbr}
\setcoords(-325,-555,0)(-900,-2100,0)[0,-2400]
\particleadjusty=-40
\or % \FLIPPEDLONG
\setcoords(-505,-275,0)(-900,-2100,0)[0,-2400]
\gdef\upperunitbox{\Lbr}   \gdef\lowerunitbox{\Rbr}
\particleadjusty=-30  % Lengths are in centipoints.
\else \UNIMPERROR
\fi
}
\gdef\SWphoton{  
\ifcase\LINECONFIGURATION  %\REG case
\setcoords(-825,-825,0)(0,-1250,0)[-1250,-1250]     
\SETUNITBOX(1250)[br][tl]
\or % \FLIPPED case
\setcoords(-175,-1425,0)(-625,-625,0)[-1250,-1250]  
\SETUNITBOX(1250)[tl][br]
\or % \CURLY case.
\setcoords(-900,-900,0)(0,-1410,0)[-1400,-1400]     
\SETUNITBOX(1400)[br][tl]
\or % \FLIPPEDCURLY case
\setcoords(-200,-1600,0)(-700,-700,0)[-1400,-1400]  
\SETUNITBOX(1400)[tl][br]
\or % \FLAT case
\setcoords(-800,-800,0)(0,-1200,0)[-1200,-1200]     
\SETUNITBOX(1200)[br][tl]
\or % \FLIPPEDFLAT case
\setcoords(-200,-1400,0)(-600,-600,0)[-1200,-1200]  
\SETUNITBOX(1200)[tl][br]
\else \UNIMPERROR
\fi
}
\gdef\Wphoton{
\ifcase\LINECONFIGURATION %\REG case
\setcoords(-2245,-1245,0)(-150,-400,0)[-2005,0]
\gdef\upperunitbox{\Frown}   \gdef\lowerunitbox{\Smile}
\or % \FLIPPED case
\setcoords(-2245,-1245,0)(-400,-150,0)[-2005,0]
\gdef\upperunitbox{\Smile}   \gdef\lowerunitbox{\Frown}
\else \UNIMPERROR
\fi
\particleadjustx=57 % Lengths are in centipoints.
\ifnum\numwigglespo=\number\numwiggles \particleadjustx=0  \fi
\numlowerunits=\numwigglespo   \numupperunits=\numwiggles
}
\gdef\NWphoton{  
\ifcase\LINECONFIGURATION  %\REG case
\setcoords(-200,-1425,0)(625,625,0)[-1250,1250]   
\SETUNITBOX(1250)[bl][tr]
\or % \FLIPPED case
\setcoords(-825,-825,0)(0,1250,0)[-1250,1250]     
\SETUNITBOX(1250)[tr][bl]
\ifnum\numwigglespo > \number \numwiggles \particleadjusty=-15 \fi
\or % \CURLY case.
\setcoords(-200,-1600,0)(700,700,0)[-1400,1400]   
\SETUNITBOX(1400)[bl][tr]
\or % \FLIPPEDCURLY case
\setcoords(-900,-900,0)(0,1400,0)[-1400,1400]     
\SETUNITBOX(1400)[tr][bl]
\or % \FLAT case.
\setcoords(-200,-1400,0)(600,600,0)[-1200,1200]   
\SETUNITBOX(1200)[bl][tr]
\or % \FLIPPEDFLAT case
\setcoords(-800,-800,0)(0,1200,0)[-1200,1200]    
 \SETUNITBOX(1200)[tr][bl]
\else \UNIMPERROR
\fi
}

\fi
\selectphoton
}
\gdef\selectgluon{   
\ifnum\gluoncount=0
\fi
\selectgluon
}
\gdef\selectspecial{\UNIMPERROR}
\gdef\checkvertex{ %immediately re-defines
\ifnum\vertexcount=-1
\fi}
\gdef\drawvertex#1[#2#3](#4,#5)[#6]{\checkvertex\drawvertex#1
[#2#3](#4,#5)[#6]}
\gdef\vertexcap#1{\checkvertex\vertexcap#1}
\gdef\vertexcaps{\checkvertex\vertexcaps}
\gdef\vertexlink#1{\checkvertex\vertexlink#1}
\gdef\vertexlinks{\checkvertex\vertexlinks}
\gdef\stemvertex#1{\checkvertex\stemvertex#1}
\gdef\stemvertices{\checkvertex\stemvertices}
\gdef\flipvertex{\checkvertex\flipvertex}
\global\arrowlength=349  % Length of arrow
\gdef\drawarrow[#1#2](#3,#4){
\global\LDIR=#1
\SETDIR
\global\boxlengthx=#3  
\global\boxlengthy=#4  % The Arrow co-ordinates.
\ifdim#2=1pt  
\adjx=\arrowlength      \adjy=\arrowlength
\multiply\adjx by \XDIR \multiply\adjy by \YDIR  % Set in \SETDIR
\slanttest(\adjx,\adjy)
\global\advance\boxlengthx by \adjx    \global\advance\boxlengthy by \adjy
\fi
\ifnum\phantomswitch=0\put(\boxlengthx,\boxlengthy)
{\vector(\XDIR,\YDIR){0}}\fi
}  % END OF \drawarrow.
\gdef\SETFRONTSTEM{
\EITHERSTEM=\FRONTSTEM   \advance\EITHERSTEM by \BACKSTEM
\ifdim\EITHERSTEM>0pt
\global\stemlengthx=\stemlength   \global\stemlengthy=\stemlength
\global\absstemlength=\stemlength
\SETDIR
\gslanttest(\stemlengthx,\stemlengthy)
\gslanttest(\absstemlength,\REG)  
\ifnum\XDIR=0 \stemlengthx=0 \fi
\ifnum\YDIR=0 \stemlengthy=0 \fi
\global\multiply\stemlengthx by \XDIR
\global\multiply\stemlengthy by \YDIR
\ifdim\FRONTSTEM=1pt
\ifnum\phantomswitch=0
          \put(\pfrontx,\pfronty){\line(\XDIR,\YDIR){\absstemlength}}\fi
\global\advance\plengthx by \stemlengthx
\global\advance\plengthy by \stemlengthy
\global\advance\PFRONTx by \stemlengthx
\global\advance\PFRONTy by \stemlengthy
\global\advance\pmidx by \stemlengthx
\global\advance\pmidy by \stemlengthy
\global\advance\pbackx by \stemlengthx
\global\advance\pbacky by \stemlengthy
\ifnum\LTYPE=3
\global\photonfrontx=\PFRONTx  \global\photonfronty=\PFRONTy
\global\photonbackx=\pbackx    \global\photonbacky=\pbacky
\fi  % END LTYPE
\ifnum\LTYPE=4
\global\gluonfrontx=\PFRONTx  \global\gluonfronty=\PFRONTy
\global\gluonbackx=\pbackx    \global\gluonbacky=\pbacky
\fi  % END LTYPE
\fi  % END FRONTSTEM
\fi  % END EITHERSTEM
}    % end of \SETFRONTSTEM
\gdef\SETBACKSTEM{
\ifdim\BACKSTEM=1pt
\ifnum\phantomswitch=0
       \put(\pbackx,\pbacky){\line(\XDIR,\YDIR){\absstemlength}}\fi
\global\advance\plengthx by \stemlengthx
\global\advance\plengthy by \stemlengthy
\global\advance\pbackx by \stemlengthx
\global\advance\pbacky by \stemlengthy
\fi  % END BACKSTEM
\global\stemlength=275  \FRONTSTEM=0pt  \BACKSTEM=0pt 
}    % END OF \SETBACKSTEM
\gdef\drawloop#1[#2#3](#4,#5){  %RECURSIVE.

\global\newcount\loopfrontx    \global\newcount\loopfronty
\global\newcount\loopbackx    \global\newcount\loopbacky
\global\newcount\loopmidx    \global\newcount\loopmidy
\global\newdimen\CENTRALLOOP
\gdef\drawloop#1[#2#3](#4,#5){
\global\CENTRALLOOP=0pt  % non-central is default
\global\LINETYPE=#1
\ifnum\LTYPE=\gluon\relax\else\UNIMPERROR\LTYPE=1
\message{Reverting to Gluons}
\fi
\global\LINEDIRECTION=#2  %initial loop direction
\global\fourthlineadjx=#3 %number of eighths of loop
\ifnum\fourthlineadjx=0 % (x,y) now midpoint.
  \global\CENTRALLOOP=1pt  % non-central is default
  \global\fourthlineadjx=8
  \global\LDIR=0
\fi
\global\fourthlineadjy=\fourthlineadjx  % a conveniently unused variable.
\global\advance\fourthlineadjy by -4
\global\loopfrontx=#4   \global\loopfronty=#5
\ifdim\CENTRALLOOP=1pt
  \global\advance\loopfrontx by -2413  \global\advance\loopfronty by -425
\fi                          % diameter of gluonloop is 4825 to 4830 cpt.
\global\unitboxnumber=1  % default; \gluoncase
\ifnum\LINETYPE=\photon \unitboxnumber=2 \fi
\checkdir
\drawline\LINETYPE[\LDIR\LCONFIG](\loopfrontx,\loopfronty)
[\unitboxnumber]
\DRAWLOOP
\ifnum\fourthlineadjy>-1 % at least 1/2 a loop
\global\loopmidx=\loopfrontx   \global\loopmidy=\loopfronty
\global\advance\loopmidx by \loopbackx  \global\advance\loopmidy by 
\loopbacky
\divide\loopmidx  by 2 \divide\loopmidy by 2  % midpoints of loop
\ifdim\CENTRALLOOP=1pt
  \global\advance\loopfrontx by 200    \global\advance\loopfronty by 425
  \global\advance\loopbackx by -200    \global\advance\loopbacky by -425
\fi
\fi % end of \ifnum\fourthlineadjy>-1
}
\gdef\DRAWLOOP{
\global\advance\fourthlineadjx by -1
\ifnum\fourthlineadjx=0\relax  % finished!
\else
\ifnum\fourthlineadjx=\fourthlineadjy % opposite side of loop
   \global\loopbackx=\pbackx   \global\loopbacky=\pbacky
\fi
\global\advance\LDIR by 1
\moduloeight\LDIR
\checkdir
\drawline\LINETYPE[\LDIR\LCONFIG](\pbackx,\pbacky)[\unitboxnumber]
\fi % end \ifnum\fourthlineadjx
\ifnum\fourthlineadjx>1 \DRAWLOOP  \fi  % recursive
}
\gdef\checkdir{
\ifnum\LTYPE=\gluon
\ifodd\LDIR \global\LCONFIG=0 \else \global\LCONFIG=2 \fi
\fi %end of \gluon
}

\drawloop#1[#2#3](#4,#5)}
\Feynmanlength  % Set length scale to centipoints.

\begin{document}

\renewcommand{\thefootnote}{\fnsymbol{footnote}}

\thispagestyle{empty}

\hfill \parbox{45mm}{
MPI-PhT/95-79 \par
LMU-TPW 95-12 \par
August 1995}

\vspace*{15mm}

\begin{center}
{\LARGE General properties of the decay amplitudes \\
for massless particles.}

\vspace{22mm}

{\large Gaetano Fiore}
\footnote{A. Von Humboldt Fellow.}
\footnote{e-mail address: fiore@lswes8.ls-wess.physik.uni-muenchen.de}

\medskip

{\em Sektion Physik der Universit\"at M\"unchen, Ls.\ Prof.\ Wess \par
Theresienstrasse 37, D 80333 M\"unchen, Germany}

\bigskip

\centerline{and}

\bigskip

{\large Giovanni Modanese$^*$}
\footnote{e-mail address: modanese@science.unitn.it}

\medskip

{\em Max-Planck-Institut f\"ur Physik \par
Werner-Heisenberg-Institut \par
F\"ohringer Ring 6, D 80805 M\"unchen, Germany}

\medskip

\end{center}

\vspace*{10mm}

\begin{abstract}

 We derive the kinematical constraints which characterize the decay
 of any massless particle in flat spacetime. We show that in
 perturbation theory the decay probabilities of photons and
 Yang-Mills bosons vanish to all orders;  the decay probability
of the graviton vanishes to one-loop order for graviton loops and to
all orders for matter loops.
 A general power counting argument indicates in which conditions a 
 decay of a massless particle could be possible: the lagrangian 
 should contain a self-coupling without derivatives and with a 
 coupling constant of positive mass dimension.

 \bigskip \bigskip
 
 \end{abstract}
 
  The massless particle which we best know, the photon, is certainly
 stable for very long periods. The experimental evidence concerning 
 the properties of the neutrino (admitted it is really massless) is 
 less strong, but it is generally regarded as stable too.
 
 Nevertheless, kinematics allows in principle the decay of a massless
 particle, provided the products are massless and their momenta
 have the same direction and versus of the initial momentum 
 (compare Section 1). This means that the Mandelstam variables of 
 the process vanish, so that its amplitude, regarded as a function 
 of Mandelstam variables, must be computed in this particular limit
 \footnote{In the four-particle amplitude we mean by Mandelstam variables
 the usual ones, $s$, $t$, $u$; for amplitudes with more
 external massless particles, they are taken to be all the possible
 scalar products between the external four-momenta.}.
 Moreover, even if the limit of the amplitude is not zero, the phase
 space for the products reduces to a line in momentum space and 
 therefore its volume tends to vanish.
 
 An almost exact ``collinearity'' of the products is usually
 observed in the decay of any particle for which $E \gg m
 $. Four-momentum conservation implies (see for instance
 \cite{synge}) that the mass of a particle produced in an
 annihilation process is proportional to the sine of the angle
 between the momenta of the colliding particles; conversely,
 in the limit $m \to 0$ the decay produces collinear particles.
 
 Limits of this kind ($m \to 0$) are common in the treatment
 of infrared singularities in quantum field theory (\cite{ir}; see 
 also Sect.\ 5). In our case, however, the assignment of an 
 infinitesimal mass $m$ to the particles involved in the decay 
 is unsuitable as a regularization technique.
 In fact, let us consider the decay of a massless
 particle of energy $E$ into $n$ collinear particles of the
 same kind, with energies $E_i$ such that $\sum_{i=1...n}
 E_i=E$. This process is kinematically allowed (compare
 Section 1; $n$ must be odd if the initial particle has
 nonzero helicity), but if we give the particles an infinitesimal
 mass, it becomes obviously impossible (suppose to observe
 it in the rest system of the initial particle).
 
 We shall then work from the beginning with massless particles
 and introduce a different regularization, involving a weak
 external source $J$ which gives the initial particle an infinitesimal
 additional energy (and/or momentum) $\omega$, so that its 4-momentum 
 is put slightly off-shell. This regularization technique
 proves to be quite effective, as it also allows an estimate
 of the decay probability by power counting.

Let us now come to the specific cases we treated.
In QED it is possible to show in a general way by means of
the Ward identities that the decay amplitude for $\gamma \to
\gamma_1 + ... + \gamma_n$ ($n$ odd)
is a symmetrized sum of terms which 
can be factorized into a 
finite scalar part and a tensor part that vanishes when all
the external momenta are aligned.
An analogous reasoning holds for
the neutrino. In both cases, it is crucial that the loop amplitudes
contain in the denominator the masses of the fermions or of the vector
bosons, respectively.

Another example of massless particle is the graviton. Here we do not 
have any experimental evidence yet. It has been suggested \cite{efr}
that the non-linearity of Einstein equations could lead to
a ``frequency degeneration'' in gravitational waves, a phenomenon
which from the quantum point of view would correspond to a decay
of the graviton into more gravitons of smaller energy. We were able
however to prove through a generalization of the procedure applied
to QED that the amplitude of this process vanishes in 
perturbation theory around the 
flat background. In this case the negative 
mass dimensionality of the Newton constant plays a role
analogous to the fermion masses in QED. At the non perturbative
level, the hypothesized existence of a small scale cosmological 
constant could change the situation (see below).

The case of the gluon, although physically quite academic due to
the confinement, is particularly interesting because the amplitude
of the decay $g \to g_1 + ... + g_n$ ($n$ odd) is finite for
$n=3$ and divergent for $n \geq 5$. (The Ward identities still allow
a factorization of this amplitude like in QED, 
but the scalar parts now contain
poles.) Nevertheless, the total decay probability is zero because the
phase space for the products is suppressed strongly enough to
compensate for the divergence in the amplitude. We thus have here
a typical example of cancellation of infrared divergences in the
computation of a physical quantity.

A general power counting argument indicates
in which conditions a real decay of a massless particle
could be possible: the lagrangian should contain a self-coupling without 
derivatives and with a coupling constant of positive mass dimension.
This is precisely what happens in quantum gravity in the presence
of a cosmological constant, and in fact it has been suggested that in
this theory strong infrared effects could become relevant \cite{woo}.
But one must remind that in the lagrangian 
the cosmological constant also
multiplies a term which is quadratic in the field and
thus generates an effective mass for the graviton (if $\Lambda < 0$)
or an unstable theory (if $\Lambda > 0$) \cite{vel}. A possible way to elude
the problem is to admit, like in lattice theory, that the effective
cosmological constant vanishes on large scales but not on small
scales and is negative in sign (compare Section 5). This latter approach is 
however out of the scope of our paper.

The structure of the article is the following.
Section 1 is concerned with kinematics.
In Section 1.1 we give
a list of simple kinematical properties which characterize the
decay of any massless particle. These properties are only due to
Lorentz invariance and to the conservation of the total four-momentum
and angular momentum. In Section 1.2 we
reexpress in a more manageable form the Lorentz-invariant 
decay measure defined
on the phase space of $n$ massless product particles, under the condition
that also the initial particle is massless;
specializing to the case $n=2$ we compute explicitly the lowest order 
decay probability in the 
toy-model scalar $\lambda \phi^3$ theory. In Section 1.3
we introduce an infrared regularization which allows the computation
of the decay amplitudes in the limit of vanishing Mandelstam variables.
In Section 2 we give a dimensional estimate of the decay
probability of the photon, the neutrino, the gluon and the graviton.
After recalling in Section 3 how the exact proper vertices are
connected to the complete perturbative expression for the decay amplitude,
in Section 4 we use the Ward identities for QED, Yang-Mills theory
(YM) and Einstein quantum gravity (QG) to give an estimate of the
regularized amplitudes. In Section 5 we comment on the relation 
between the infrared singularities which occur in our computations 
and the usual infrared singularities of quantum field theory.
Finally we present a few brief speculations about the
possible role of a non-vanishing cosmological constant in the
decay of the graviton.

\section{General kinematic properties.}

\subsection{Consequences of Lorentz invariance.}

We list here the most general properties of the decay
of a massless particle. They are due only to the Lorentz invariance
of the process and to the conservations of the total four-momentum
and angular momentum.
As we mentioned in the introduction, some of them can be proven
taking the limit $m \to 0$ in the corresponding formulas for
massive particles \cite{synge}. Properties 1, 2, 3, 6 can also 
be found in ref.\ \cite{mod}.

\medskip
\noindent {\bf Property 1.} -- A massless particle can only decay into
massless particles. -- In fact, through a suitable Lorentz boost we
can make the energy of the initial state arbitrarily small. If, {\it per
absurdum}, in the final state massive particles were present, the
energy of this state would be in any reference frame equal or
bigger than the sum of the masses.

\medskip
\noindent {\bf Property 2.} -- Let us suppose that the impulse
$\vec{p}^{\ 0}$ of the initial particle is oriented in a certain
direction and versus, for instance let its four-momentum have the form
\begin{equation}
  p^0 = (E^0,\, E^0,\, 0,\, 0)
\end{equation}
Then also the impulses $\vec{p}^{\ 1} ... \vec{p}^{\ n}$ of the
$n$ product particles are oriented in the same direction and versus;
in our example we shall have (Fig.\ 1)
\begin{equation}
  p^i = (E^i,\, E^i,\, 0,\, 0); \qquad i=1,...,n; \qquad
  \sum_{i=1}^n E^i = E^0.
\end{equation}
In an arbitrary Lorentz frame this can be rewritten as
\begin{equation}
  p^i = \lambda_i p^0, 
\label{pippo}
\end{equation}
where $1>\lambda_i>0$, $i=1,2,...,n$, $\sum_i\lambda_i=1$.

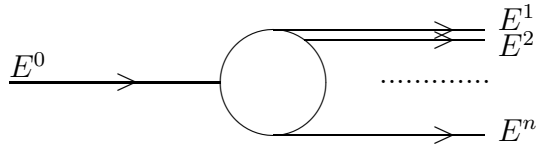
\begin{figure}[hh]
\begin{center}
\begin{picture}(30000,6000)
\drawline\fermion[\E\REG](0,3000)[8000]
\put(0,3200){$E^0$}
\put(4000,2700){$>$}
\put(10000,3000){\circle{4000}}
\drawline\fermion[\E\REG](10000,5000)[8000]
\put(18500,5000){$E^1$}
\put(16000,4700){$>$}
\drawline\fermion[\E\REG](11200,4600)[6800]
\put(18500,4000){$E^2$}
\put(16000,4300){$>$}
\drawline\fermion[\E\REG](10000,1000)[8000]
\put(16000,700){$>$}
\put(18500,800){$E^n$}
\put(14000,3000){.............}
\end{picture}
\end{center}
\caption{\label{colli} Collinearity property (Property 2).}
\end{figure}

-- Also this property depends on the fact that through a suitable
Lorentz boost along $z$ we can make the energy of the initial state
arbitrarily small; while if {\it per absurdum} in the final state some 
transversal momenta were present, their contribution to the energy would 
not be affected by the boost.

\medskip
\noindent {\bf Property 3.} -- If the initial particle has helicity
$h$ and decays into $n$ particles of the same helicity, $n$ must
be odd. -- The proof follows directly from Property 2 and from the
conservation of the angular momentum.

\medskip
\noindent {\bf Property 4.} -- In the decay of a massless particle, all the
scalar products $(p^i \cdot p^j), \ i,j=0,1,...,n$ vanish.
This means that the Mandelstam variables vanish. -- The proof follows
directly from Property 2.

\medskip
\noindent {\bf Property 5.} -- If $\varepsilon^i$ represents the polarization
vector of the $i$-th particle involved in the decay, in a gauge such
that $(p^i \cdot \varepsilon^i) = 0$, then we have also
$(p^i \cdot \varepsilon^j) = 0$ for $i,j=0,1,...,n$. -- 
Once more, the proof follows directly from Property 2.

\medskip
\noindent {\bf Property 6.} -- If a massless particle decays, its
lifetime $\tau$ in a reference frame where its energy is $E^0$ has the
form
\begin{equation}
  \tau = \xi E^0
\label{bella}
\end{equation}
where $\xi$ is a constant which depends on the dynamics of the process
and has dimension $[mass]^{-2}$. -- This property holds also for
massive particles, for which the constant takes the form $\xi= 
\tau_{rest}/m$. The proof is elementary (see for instance \cite{mod}).

\subsection{The decay phase space measure $d\mu_n$.}

 We recall that according to quantum field theory the decay probability
 (per unit time) should be computed by the general formula
 \begin{equation}
   \tau^{-1}=\Gamma = \frac{1}{2E^0} \sum_{n \geq 2}
   \int \prod_{i=1}^n \frac{d^3 p^i}{(2 \pi)^3 2E^i}
   \, \delta^4 \left( p^0 - \sum_{i=1}^n p^i \right) |T_n|^2
 \label{probability}
 \end{equation}
 where $T_n$ is the quantum amplitude for the decay process into $n$ 
product
 particles of momenta $\{p^i\}$. If the final particles have helicity or 
 internal quantum symmetry numbers $T_n$ includes the sum over these 
 degrees of freedom.
 
 Actually, both eq.s (\ref{bella}) and (\ref{probability}) give
 physically realistic predictions as far as:
 
 \noindent (1) the energy uncertainty $\Delta E$ of the first particle
 fulfils the condition $\Delta E\ll E$; 
 
 \noindent (2)\footnote{We thank M. Abud
for drawing our attention to this point and to several further subtleties 
which are required by a correct
physical interpretation of infrared divergences related to massless 
particles}
 the finite energy resolution $\epsilon$ of the decay
 detector can be neglected. In general the detector will be unable to 
 recognize a decay process in which one of the outcoming particles has 
 energy $E'$ such that $E-E'\ll \epsilon$. In order to compute the 
 correct detection probability $\Gamma_{\epsilon}$ one should in principle
 subtract from formula (\ref{probability}) the total probability of all
 events of this kind. Nevertheless, for the theories considered in this 
 paper one finds that this effect is indeed negligible (in perturbative 
 QED, YM and QG we will find $\Gamma=0$,  whence it follows
 $\Gamma_{\epsilon}=0$, since $\Gamma_{\epsilon}\le\Gamma$; in the
 $\lambda\phi^3$ toy-model at order $O(\lambda^2)$ considered below
 one finds that $\Gamma-\Gamma_{\epsilon}\sim\epsilon$). We plan to
 devote more attention to the general issue elsewhere,
 by considering examples of theories for which condition (2) is not 
 fulfilled. This requires an approach to IR divergences as in the
 Kinoshita-Lee-Nauenberg theorem \cite{ir}.
 
 \medskip
 A closer look at the measure appearing in the integrals on the RHS of
 formula (\ref{probability}) is now very useful. When all particles are 
 massless, it is possible to express the Lorentz-invariant decay measure
 \begin{equation}
   d\mu_n = \prod_{i=1}^n \frac{d^3 p^i}{(2 \pi)^3 2E^i}
   \, \delta^4 \left( p^0 - \sum_{i=1}^n p^i \right)
 \label{misura}
 \end{equation}
 in the following form:
 \begin{eqnarray}
   d\mu_n & = & \frac{2\alpha_{2n-2} }{E^0} \left[ \prod_{i=1}^n 
   d^3 p^i \, \delta^2(\vec{p}^{\ i}_T) \, \theta(p^{\ i}_L)
   \right] \delta \left( \vec{p}^{\ 0} - \sum_{i=1}^n \vec{p}^{\ i}_L
   \right) \nonumber \\
   & & \times \left[ 2 \left( E^0 - \sum_{i=1}^n |\vec{p}^{\ i}|
   \right) - \frac{1}{E^0} \left( \sum_{i=1}^n \vec{p}^{\ i}_T
   \right)^2 \right]^{n-2} ,
 \label{capricciosa}
 \end{eqnarray}
 where $\vec{p}^{\ i}_L$ and $\vec{p}^{\ i}_T$ denote the longitudinal
 and trasversal part of the impulse $\vec{p}^{\ i}$ with respect to the
 direction and versus identified by $\vec{p}^{\ 0}$, $\theta$ is the
 step function, and the
 adimensional coefficients $\alpha$ are those which appear in the 
 expression of $\delta^m(\vec{x})$ in polar coordinates:
 \begin{equation}
   \delta^m(\vec{x}) = \alpha_m^{-1} \delta(|\vec{x}|^2)
   |\vec{x}|^{2-m} .
 \end{equation}
 The $\delta$-functions occurring in formula (\ref{capricciosa})
 show that the support of $d\mu_n$ is concentrated around (the
 infinitesimal neighbourhood of) the collinearity region, 
 which is characterized by all sets $\{p^i\}$ satisfying relation 
 (\ref{pippo}). 
 
 The collinearity property (\ref{pippo}) follows from the sole condition
 $p^0 - \sum_{i=1}^n p^i=0$ [imposed by the $\delta$-function
 contained in formula (\ref{misura})] if all $p^i$ are null vectors.
 In fact, we observe that
 \begin{equation}
   \left( \sum_{i=1}^n |\vec{p}^{\ i}| \right)^2 -
   \left| \sum_{i=1}^n \vec{p}^{\ i} \right|^2 =
   \left( \sum_{i=1}^n p^i \right)^2 =(p^0)^2= 0 .
 \label{ciao}
 \end{equation}
 The 3-vector $\sum_{i=1}^n \vec{p}^{\ i}$ has length
 $\ell \leq \sum_{i=1}^n |\vec{p}^{\ i}|$, and the equality holds only
 if $\vec{p}^{\ i} = \lambda_i \vec{p}$, for some $\vec{p}$ and some
 $\{\lambda_i\}$ all of the same sign; inserting this into the relation
 $p^0 - \sum_{i=1}^n p^i=0$ we find eq.\ (\ref{pippo}).
 
 \medskip
 The squared amplitude $|T_n|^2$ depends only on the Lorentz 
 invariants $(p^i\cdot p^j)$. But in the collinearity region 
 $p^i\cdot p^j=0$. Thus in this region a finite $|T_n|^2$ may only be 
 a function (possibly trivial) of the invariants $\lambda_i$
 defined in formula (\ref{pippo}); in this case the corresponding integral 
 can be easily performed and gives a finite (possibly vanishing) result.
 In particular, when $n \geq 3$ the factor in the square bracket of 
 formula (\ref{capricciosa}) is set equals to zero by the 
 $\delta$-functions, and thus if the amplitude of the decay is finite, 
 the corresponding total probability is zero.
 If $|T_n^2|$ diverges, we may introduce a suitable regularization
 in order to make the integration easier (see Section 1.3). 
 
 For a massless scalar field theory
 with self-coupling of the form $\lambda \phi^3$ the phase space
 integral (\ref{capricciosa}) with $n=2$ coincides, up to a
 factor $\lambda^2$, with the probability of the decay of a
 particle into two particles, computed perturbatively to lowest
 order. This is a concrete example of computation of a finite
 decay probability, although with the known limitations of
 the $\lambda \phi^3$ theory
 \footnote{It is known that the action is not limited from below and 
 that the radiative corrections do not preserve $m=0$.}.
 
 Setting $n=2$ in (\ref{capricciosa}) and performing the integral
 (the square amplitude does not depend on $p^1, \ p^2$ and is equal
 to $\lambda^2$) we obtain
 \begin{equation}
   \int d \mu_2 = \frac{2\alpha_2}{E^0} \int dp^1 dp^2 \,
   \delta[E^0-(p^1)^1-(p^2)^1] \theta[(p^1)^1] \theta[(p^2)^1] =
   2 \alpha_2 .
 \end{equation}
 
 The present conclusion that $\Gamma$ is finite to order $\lambda^2$
 coincides with that of the dimensional analysis (considered in 
 Section \ref{power}) applied to this case, in which the coupling 
 constant has positive mass dimension.

 \subsection{Regularization through an external source.}
 \label{regu}
 
 We would like now to introduce an infrared regularization in order
 to allow a quick estimate of the integrals (\ref{probability}) in all
 cases (including the case in which the amplitude $T_n$ diverges
 on the collinearity region).
 
 Obtaining such a regularization is not trivial. The most common infrared
 regularization technique, which consists in giving the soft particles a small
 mass $\mu$ which eventually goes to zero, does not work in the present 
case,
 because the (regularized) process in which one particle of mass $\mu$
 decays into more particles of the same mass has obviously zero 
probability.
 
 \begin{figure}[hh]
 \begin{center}
 \begin{picture}(30000,10000)
 \drawline\fermion[\E\REG](0,3000)[8000]
 \put(3000,1000){I}
 \put(4000,2700){$>$}
 \put(8000,3000){\circle*{400}}
 \drawline\fermion[\E\REG](\fermionbackx,\fermionbacky)[5000]
 \put(10000,1000){II}
 \put(11000,2700){$>$}
 \drawline\fermion[\N\REG](8000,3000)[5000]
 \drawline\fermion[\NE\REG](7400,8300)[1800]
 \drawline\fermion[\NW\REG](8600,8300)[1800]
 \put(8000,9000){\circle{2000}}
 \put(14000,3000){\circle{2000}}
 \drawline\fermion[\E\REG](14000,4000)[8000]
 \put(20000,3300){$>$}
 \drawline\fermion[\E\REG](14800,3600)[7400]
 \put(21000,3700){$>$}
 \drawline\fermion[\E\REG](14000,2000)[8000]
 \put(20000,1700){$>$}
 \put(16000,3000){.............}
 \put(17000,1000){III}
 \end{picture}
 \end{center}
 \caption{\label{factor} Factorization of the decay amplitude.}
 \end{figure}
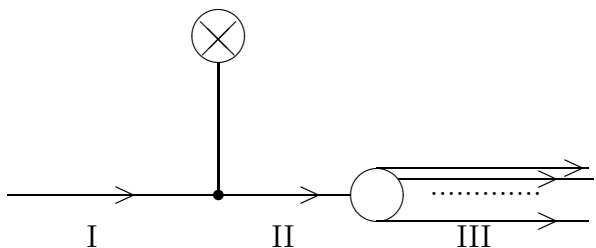
 
 Instead, a better approach is to put external momenta slightly off-shell in
 a way controlled by an infinitesimal parameter $\omega$.
                                                       
 Let us suppose (Fig.\ 2) that a very weak external source
 $J$ gives the decaying particle (state I) an infinitesimal additional
 energy $\omega$.
 The exact nature of the source and of the particle which
 carries the energy $\omega$ are not essential. For instance,
 if $J$ represents a classical field, the energy can be carried by an 
 on-shell boson with four-momentum $(\omega, \, 0, \, 0, \, \omega)$;
 by absorbing the boson, the initial particle gains a small transversal
 impulse (state II) too. Alternatively, the energy $\omega $ could be carried
 by an off-shell boson produced in $J$ through an annihilation process,
 with four-momentum $(\omega, \, 0, \, 0, \, 0)$; by absorbing the
 boson, the initial particle gets off shell too.
 More generally, we will assume that
 after the interaction the four-momentum of the initial particle
 will have the form
 \begin{equation}
 p^0 \prime = \hat{p^0} + \omega b^0,
 \end{equation}
 where $b^0$ is adimensional and $(\hat{p^0} )^2=0$. It is not
 necessary to make any special assumption on the four-vector $b^0$
 at this stage
 (in Sections 3, 4 we will prefer to specialize the discussion by assuming
 that $b^0\cdot p^0=0$, and other similar conditions for the $p^i$'s).
 
 At this point the decay takes place; the products (state III) have now
 a small tranversal impulse of order $\omega $ and the Mandelstam
 variables $(p^i \cdot p^j)$ are of order $\omega$
 (at least).
 The partial decay
 probability into $n$ product particles is written as a sum over intermediate
 states (compare (eq.\ \ref{probability})
 \begin{equation}
   \Gamma_n = \lim_{\omega \to 0}  \frac{1}{2E^0_{\omega}}
   \int \prod_{i=1}^n \frac{d^3 p^i}{(2 \pi)^3 2E^i}
   \, \delta^4 \left( p^{II}_{\omega } - \sum_{i=1}^n p^i \right)
   | \langle II_{\omega } | {\cal T} | III \rangle |^2 
 \label{verde}
 \end{equation}
 where ${\cal T}$ is the appropriate evolution operator.
 When $\omega \to 0$, the factor
 $1/E^0_{\omega}$ tends to $1/E^0$,
 which is the dependence that
 we expect on the basis of Lorentz invariance (compare Property 6).
 Thus in this limit the integral $I_n$ appearing in
 the preceding formula does not depend on $E$.
 Summing up we obtain
 \begin{equation}
   \Gamma_n =\frac{1}{2E^0}\lim_{\omega \to 0} I_n(\omega);
 \label{verdissima}
   \end{equation}                                                                  the only
 the only
 massive parameters on which $I_n(\omega)$ depends are $\omega$
 and the massive parameters possibly present in the theory that we
 are considering. This allows in most cases to estimate dimensionally
 whether $\Gamma_n$ is finite, vanishes or diverges in the limit
 $\omega \to 0$
 [note that the mass dimensions of $I_n$,
 $| \langle II_{\omega '}|{\cal T} | III \rangle |^2$ and
 of $d\mu_n$ are respectively equal to $2,2(3-n),{2n-4}$].
 We shall give some examples of this in the next Section.

 \section{Power counting.}
 \label{power}
 
 In several cases the integral $I_n$ can be estimated by simple 
 arguments (often dimensional considerations alone are enough).
 
 For instance,
 in QED the four-photons amplitude is given to lowest order by the
 four fermions loop (fig.\ 3a). It is easy to realize that the loop integral
 gives a 4-th degree homogeneous polynomial in the dimensionless 
variables
 $\frac{p^i}{m_f}$ \cite{itz}, where $m_f$ is the mass
 of the fermion. The integral $I_3$ will therefore be proportional to
 \begin{equation}
 I_3  \sim \alpha^4 \left( \frac{\omega}{m_f} \right)^8\omega^2
 \end{equation}
 where $\alpha$ is the fine structure constant. All behaves as though
 \begin{equation}
   T_3 \sim \alpha^2 \left( \frac{\omega}{m_f} \right)^4.
 \label{pioggia}
 \end{equation}
 To be precise, the behaviour $T_3\sim \omega^4$ holds only for some 
specific
 choices of the "slightly off-shell" external momenta
 $p^i$, whereas in any case $T_3=O( \omega^2)$ at least;
 the integration transforms the remaining dependence
 of $T_3$ on $p^i$, if any, into an additional $\omega^2$ factor.
 
 The above result can be generalized to the $n$-fermions loop: the key 
point
 is that the fermionic propagators of the loop produce masses in the
 denominator. The case of the neutrino is analogous: the masses of
 $Z^0$ or $W^\pm$ appear at the denominator in the amplitude.
 In both cases, since
 the amplitude is proportional to a positive power of the regularizator
 $\omega$, it vanishes in the infrared limit due to (\ref{verde}).

 \begin{figure}[hh]
 \begin{center}
 \begin{picture}(30000,10000)
 \put(3000,3000){\circle*{400}}
 \put(7000,3000){\circle*{400}}
 \put(3000,7000){\circle*{400}}
 \put(7000,7000){\circle*{400}}
 \drawline\fermion[\N\REG](3000,3000)[4000]
 \drawline\fermion[\E\REG](3000,3000)[4000]
 \drawline\fermion[\S\REG](7000,7000)[4000]
 \drawline\fermion[\W\REG](7000,7000)[4000]
 \drawline\photon[\SW\REG](3000,3000)[4]
 \drawline\photon[\SE\REG](7000,3000)[4]
 \drawline\photon[\NE\REG](7000,7000)[4]
 \drawline\photon[\NW\REG](3000,7000)[4]
 \put(23000,3000){\circle*{400}}
 \put(27000,3000){\circle*{400}}
 \put(23000,7000){\circle*{400}}
 \put(27000,7000){\circle*{400}}
 \drawline\photon[\N\REG](23000,3000)[4]
 \drawline\photon[\E\REG](23000,3000)[4]
 \drawline\photon[\S\REG](27000,7000)[4]
 \drawline\photon[\W\REG](27000,7000)[4]
 \drawline\photon[\SW\REG](23000,3000)[4]
 \drawline\photon[\SE\REG](27000,3000)[4]
 \drawline\photon[\NE\REG](27000,7000)[4]
 \drawline\photon[\NW\REG](23000,7000)[4]
 \end{picture}
 \end{center}
 \caption{\label{box} (a) Fermions square loop. (b) Gravitons or gluons 
loop.}
 \end{figure}
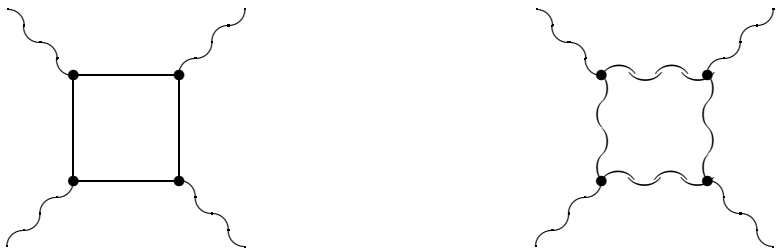
 
 In the case of pure quantum gravity we have tree and one-loop graviton
 diagrams with $k$ external legs (fig.\ 3b). Explicit expressions for the 
 $k=4$ amplitudes have been given by \cite{cho,ber}. In any case, these
 amplitudes contain
 positive powers of the constant $\kappa=\sqrt{16\pi G}$ and then, like in
 QED, they behave always like a positive power of $\omega$ and cause
 the decay probability to vanish.
 
 In the case of QCD the amplitudes do not contain dimensional constants.
 We expect that the decay amplitude of the gluon into three gluons,
 being adimensional, tends to a constant when $\omega \to 0$,
 and this is in fact what happens \cite{ber}. The decay amplitudes of a
 gluon into
 5, 7 ... gluons have mass dimensions -2, -4 ... respectively, so they
 diverge when $\omega \to 0$; but this divergence is compensated
 in the phase space integral by a bigger positive power of $\omega$
 in such a way that the probability behaves like $\omega^2/E^{(0)}$
 and thus vanishes in the limit.
 
 We are not going to apply this power counting argument to all possible
 theories and couplings, since it is in each case quite immediate.
 As a last example, we may wonder whether a photon can in principle decay
 due to the gravitational interaction, through diagrams with external
 photons and one loop of gravitons. Since the coupling constant
 $\kappa$ has mass dimension -1, while the fine structure constant
 $\alpha$ is adimensional and there are no masses involved, we conclude
 once more that the amplitude of the process vanishes in the
 infrared limit.
 
 It is clear from the discussion above that a $\Gamma_n$ different from 
 zero can be only obtained when the square amplitude is proportional to a
 sufficiently high negative power of $\omega$. Since
 in perturbation theory the coupling constants
 always appear in the numerator, this means that the amplitude must contain
 a coupling constant with positive mass dimension. We shall return
 on this point in the conclusions.
 
 \section{Diagrammatics: $\omega$-dependence of the decay amplitudes.}
 
 The dimensional arguments of the previous section determine the 
 $\omega$-dependence of the decay probability only for the pure 
 gauge theories (YM, QG), where the only parameter in the action 
 is the coupling constant. If
 additional dimensionful parameters appear in the action
 (as it happens for instance when the gauge field is coupled to
 some  massive field) the previous arguments, as we have seen in the QED
 example, must be completed by some additional information.
 In general,
 a more explicit analysis of the
 perturbative expansion and use of Feynman diagrams is  therefore needed
 in order to estimate the total decay probability.
 In this and in the following section we carry it out in such a way
 to determine not only the $\omega$-dependence of the total decay
 probability, but also of the decay amplitudes
 (i.e., of the probabilities of the single decay channels).
 The general results for the former
 will be essentially  the same as those found by
 the dimensional arguments in section 2. Thus,
 we conclude that the decay probability
 of the gauge bosons of QED, YM, QG vanish.
 
 \medskip
 Before starting, let us define a ``decay
 configuration'' as follows: it is a pair of $(n+1)$ four-momenta 
 and $(n+1)$ polarization vectors $(p^i,\varepsilon^i)_{i=0,1,...,n}$ 
 satisfying the properties $(p_i)^2=0$, $\sum_{i=0}^n p^i=0$, 
 $(\varepsilon^i \cdot p^i)=0$, $p_0^0 >0$, $p^l_0 <0$ for $l=1,...,n$.
 We thus agree that the signs of the four-momenta of the outgoing 
 particles are reversed. As we have seen, for particles with non-zero 
 helicity $n$ must be odd.
 
 We will start the analysis of the perturbative expansion
 from the tree level: a sum of  truncated connected tree-diagrams with
 $(n\!+\!1)$
 external lines will give the lowest order (in $\hbar$) contribution to the
 decay amplitude of 1 gauge boson in $n$ gauge bosons. Higher order
 corrections will involve truncated connected diagrams with one or more 
loops.
 To formally compute  the ``exact'' decay amplitude one has to replace
 in each tree diagram every  boson propagator
 with the corresponding exact
 boson propagator, and each $m$-boson vertex with
 the corresponding 
$m$-boson proper vertex (i.e. one-particle-irreducible
Green function)\footnote{In principle, propagators
and proper vertices could be computed even in two different
gauges, in order to simplify calculations, see Ref. \cite{gri}}.
 In order to get the $\hbar^r$-order approximation of
 the decay amplitude, one simply has to retain the terms of order $\le r$
 in this formal `` exact'' expression. As we will see, the
 Ward identities imply that when approaching a decay
 configuration:  (1) in QED the decay amplitude of a process
 with $m$ external photons vanishes; (2) in  QG the decay amplitude of a 
process
 with $m$ external gravitons or photons vanishes; (3) the decay amplitudes of
 processes
 with external Y.M. bosons  may be finite or diverge, but in such a way that the
 corresponding decay probabilities vanish.

\subsection{Tree level}

\begin{figure}[hh]
\begin{center}
\begin{picture}(44000,12000)
\drawline\photon[\E\REG](5000,10000)[4]
\drawline\photon[\SE\REG](\photonbackx,\photonbacky)[4]
\drawline\photon[\NE\REG](9000,10000)[4]
\put(11000,10000){$\neq 0,$}
\drawline\photon[\NW\REG](24000,10000)[4]
\drawline\photon[\SE\REG](24000,10000)[4]
\drawline\photon[\NE\REG](24000,10000)[4]
\drawline\photon[\SW\REG](24000,10000)[4]
\put(28000,10000){$\neq 0,$}
\put(38000,10000){YM}
\drawline\photon[\E\REG](0,2000)[4]
\drawline\photon[\SE\REG](\photonbackx,\photonbacky)[4]
\drawline\photon[\NE\REG](4000,2000)[4]
\put(8000,2000){$\neq 0,$}
\drawline\photon[\NW\REG](16000,2000)[4]
\drawline\photon[\SE\REG](16000,2000)[4]
\drawline\photon[\NE\REG](16000,2000)[4]
\drawline\photon[\SW\REG](16000,2000)[4]
\put(20000,2000){$\neq 0,$}
\drawline\photon[\NW\REG](28000,2000)[4]
\drawline\photon[\SE\REG](28000,2000)[4]
\drawline\photon[\W\REG](28000,2000)[4]
\drawline\photon[\NE\REG](28000,2000)[4]
\drawline\photon[\SW\REG](28000,2000)[4]
\put(32000,2000){$\neq 0,$}
\put(36000,2000){....}
\put(44000,2000){QG}
\end{picture}
\end{center}
\caption{\label{vert} Feynman vertices}
\end{figure}
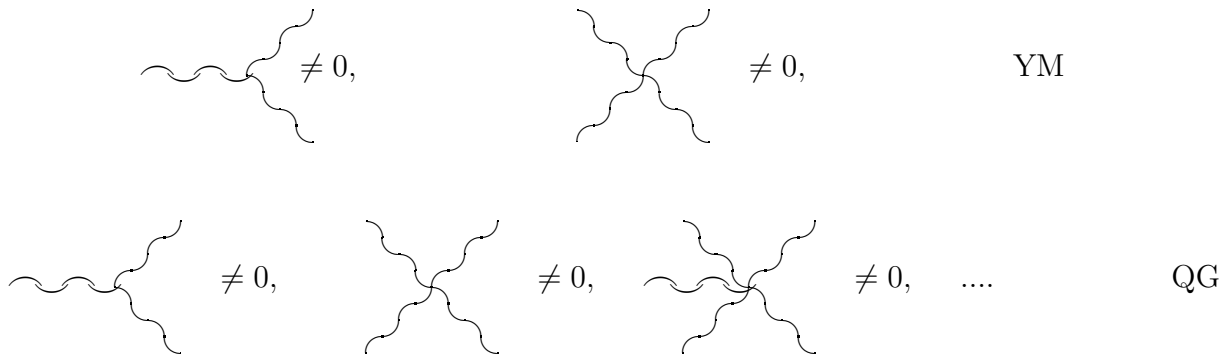

Let us start from the  Feynman vertices with $m$ gauge massless bosons
$(m\ge 3)$
[see  the actions (\ref{act1})]: we draw them in fig. (\ref{vert}). The
diagrams
are
to be understood as truncated in the external lines.
In QED there is no $m$-photon vertex.
In YM  there are only two $m$-gluon vertices (for $m=3,4$). In pure QG 
there is
one $m$-graviton vertex for every $m\ge 3$; if coupling of gravity with the
electromagnetic or the Yang-Mills fields is considered, then there are also
vertices
with $k$ spin-1 bosons
(photons or gluons) and $r$ gravitons, for $k=2,3,4$ and $r\ge 1$.
In  the figures, a wavy line  in  the QG case will denote either
a graviton or another gauge boson (a photon or a gluon).

\begin{figure}[hh]
\begin{center}
\begin{picture}(44000,34000)
\put(0,32000){$T_2^{tree}~=$}
\drawline\photon[\E\REG](6000,32000)[4]
\drawline\photon[\SE\REG](\photonbackx,\photonbacky)[4]
\drawline\photon[\NE\REG](10000,32000)[4]
\put(0,24000){$T_3^{tree}~~=$}
\drawline\photon[\NW\REG](8000,24000)[4]
\drawline\photon[\SE\REG](8000,24000)[4]
\drawline\photon[\NE\REG](8000,24000)[4]
\drawline\photon[\SW\REG](8000,24000)[4]
\put(10000,24000){+}
\drawline\photon[\E\REG](12000,24000)[4]
\drawline\photon[\SE\REG](\photonbackx,\photonbacky)[4]
\drawline\photon[\NE\REG](16000,24000)[4]
\drawline\photon[\SE\REG](\photonbackx,\photonbacky)[4]
\drawline\photon[\NE\REG](18500,26500)[4]
\put(0,16000){$T_4^{tree}~~=$}
\drawline\photon[\NW\REG](10000,16000)[4]
\drawline\photon[\SE\REG](10000,16000)[4]
\drawline\photon[\W\REG](10000,16000)[4]
\drawline\photon[\NE\REG](10000,16000)[4]
\drawline\photon[\SW\REG](10000,16000)[4]
\put(12000,16000){(QG)}
\put(16000,16000){+}
\drawline\photon[\SW\REG](19000,16000)[4]
\drawline\photon[\SE\REG](19000,16000)[4]
\drawline\photon[\NE\REG](19000,16000)[4]
\drawline\photon[\NW\REG](19000,16000)[4]
\drawline\photon[\SE\REG](21000,18000)[4]
\drawline\photon[\NE\REG](21500,18500)[4]
\put(0,8000){$T_5^{tree}~~=$}
\drawline\photon[\NW\REG](10000,8000)[4]
\drawline\photon[\SE\REG](10000,8000)[4]
\drawline\photon[\W\REG](10000,8000)[4]
\drawline\photon[\E\REG](10000,8000)[4]
\drawline\photon[\NE\REG](10000,8000)[4]
\drawline\photon[\SW\REG](10000,8000)[4]
\put(14000,8000){(QG)}
\put(18000,8000){+}
\drawline\photon[\NW\REG](23000,8000)[4]
\drawline\photon[\SE\REG](23000,8000)[4]
\drawline\photon[\E\REG](23000,8000)[4]
\drawline\photon[\NE\REG](23000,8000)[4]
\drawline\photon[\SW\REG](23000,8000)[4]
\drawline\photon[\SE\REG](27000,8000)[4]
\drawline\photon[\NE\REG](27000,8000)[4]
\put(30000,8000){(QG)}
\put(34000,8000){+}
\drawline\photon[\SW\REG](39000,8000)[4]
\drawline\photon[\SE\REG](39000,8000)[4]
\drawline\photon[\NE\REG](39000,8000)[4]
\drawline\photon[\NW\REG](39000,8000)[4]
\drawline\photon[\SE\REG](41500,10500)[4]
\drawline\photon[\NE\REG](41500,10500)[4]
\drawline\photon[\E\REG](41500,10500)[4]
\put(45000,8000){+}
\drawline\photon[\SE\REG](10000,0)[4]
\drawline\photon[\SW\REG](10000,0)[4]
\drawline\photon[\NE\REG](10000,0)[4]
\drawline\photon[\NW\REG](10000,0)[4]
\drawline\photon[\SE\REG](12500,2500)[4]
\drawline\photon[\NE\REG](12500,2500)[4]
\drawline\photon[\SE\REG](15000,0)[4]
\drawline\photon[\NE\REG](15000,0)[4]
\put(20000,0){+}
\drawline\photon[\E\REG](22000,0)[4]
\drawline\photon[\SE\REG](\photonbackx,\photonbacky)[4]
\drawline\photon[\NE\REG](26000,0)[4]
\drawline\photon[\SE\REG](\photonbackx,\photonbacky)[4]
\drawline\photon[\SE\REG](\photonbackx,\photonbacky)[4]
\drawline\photon[\E\REG](31000,0)[4]
\drawline\photon[\NE\REG](28500,2500)[4]
\drawline\photon[\SE\REG](\photonbackx,\photonbacky)[4]
\drawline\photon[\E\REG](31500,4500)[4]
\end{picture}
\end{center}
\caption{\label{tree} Tree level amplitudes: (QG) means that the diagram in
$T_5^{tree}$ is present only in QG.}
\end{figure}
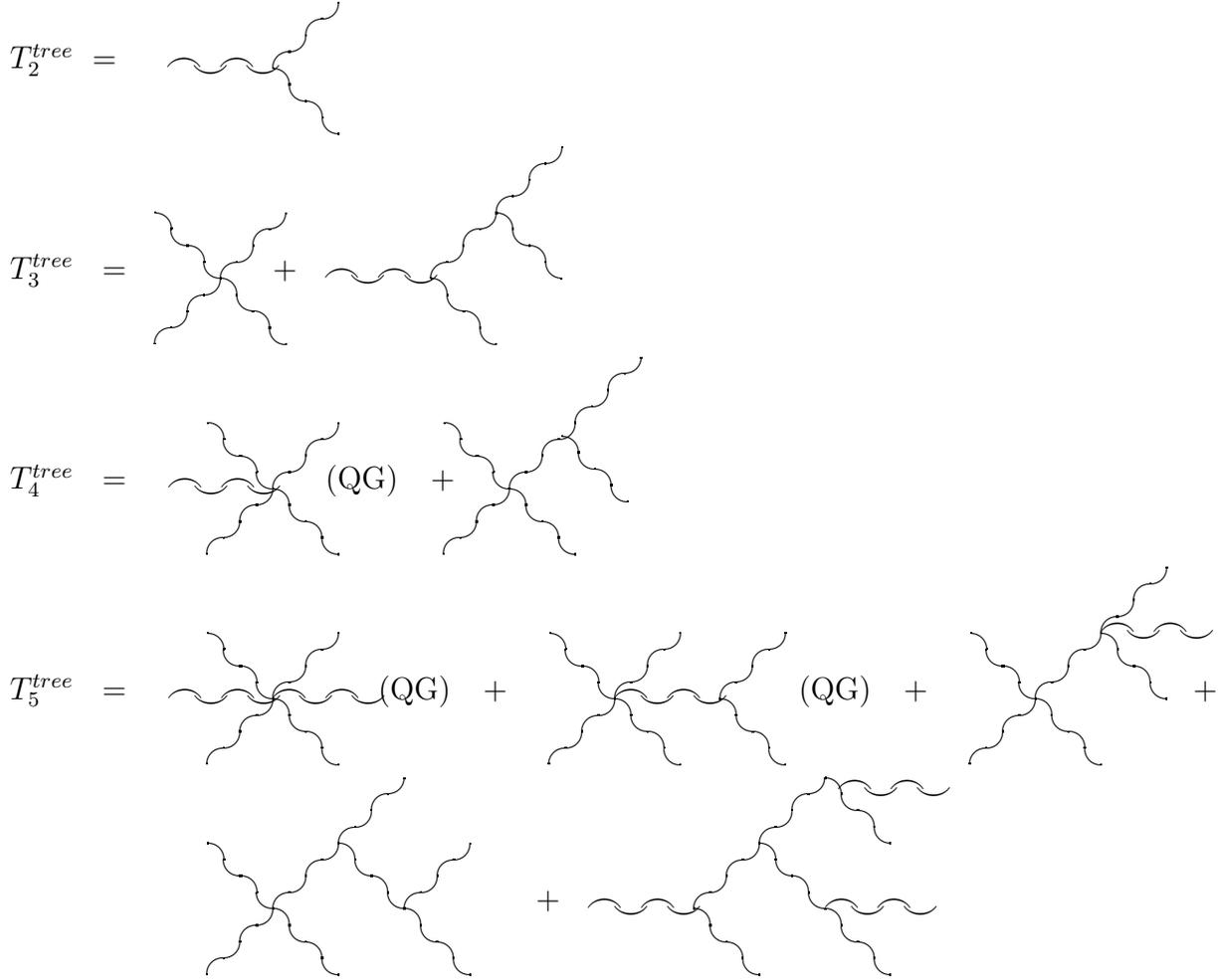

At the tree level, the decay amplitudes
$T^{tree}_2,T^{tree}_3,T^{tree}_4,T^{tree}_5,...$
of YM, QG are the sum of the diagrams in fig. (\ref{tree}).

Tree diagrams involving ghost lines do not contribute to $T_n^{tree}$. In fact,
even though ghosts are massless, diagrams with external ghosts are zero when
multiplied by physical polarization vectors, and diagrams with internal ghost
lines
(propagators) have necessarily also external ghost lines, by ghost number
conservation.
One can easily verify that in QG the decay amplitudes  with only
$m$ external gravitons or photons vanish ($T_n^{tree}=0$)
in any decay configuration, because
each vertex is quadratic in the momenta $k^i$, implying an overall $(k)^2$
dependence of each separate diagram in fig. (\ref{tree}); when contracted with
the external
polarization vectors, this will give zero, since in the decay configuration all
4-momenta
are null vectors proportional to each other.

\subsection{Higher orders}

\begin{figure}[hh]
\begin{center}
\begin{picture}(44000,12000)
\put(0,10000){$T_2~=$}
\drawline\photon[\E\REG](6000,10000)[4]
\drawline\photon[\SE\REG](\photonbackx,\photonbacky)[4]
\drawline\photon[\NE\REG](10000,10000)[4]
\put(10000,10000){\circle*{1000}}
\put(0,2000){$T_3~=$}
\put(8000,2000){\circle*{1000}}
\drawline\photon[\NW\REG](8000,2000)[4]
\drawline\photon[\SE\REG](8000,2000)[4]
\drawline\photon[\NE\REG](8000,2000)[4]
\drawline\photon[\SW\REG](8000,2000)[4]
\put(10000,2000){+}
\put(16000,2000){\circle*{1000}}
\drawline\photon[\E\REG](12000,2000)[4]
\drawline\photon[\SE\REG](\photonbackx,\photonbacky)[4]
\drawline\photon[\NE\REG](16000,2000)[4]
\drawline\photon[\SE\REG](\photonbackx,\photonbacky)[4]
\drawline\photon[\NE\REG](18500,4500)[4]
\put(18500,4500){\circle*{1000}}
\end{picture}
\end{center}
\caption{\label{trunc} Exact amplitudes}
\end{figure}
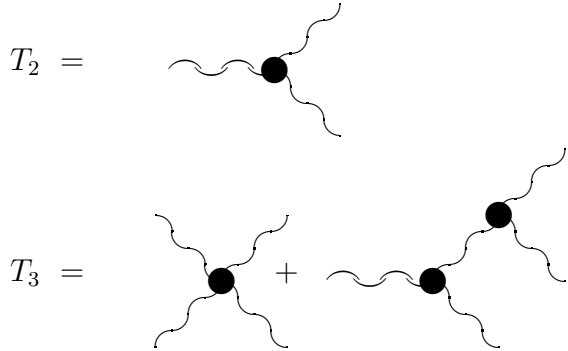

To formally compute  the ``exact'' decay amplitude one has to replace
in each tree diagram every  boson propagator with the corresponding exact
boson propagator, and each $m$-boson vertex with
the corresponding $m$-boson proper vertex (i.e. one-particle-irreducible
Green function), as depicted in fig. (\ref{trunc}); there we have symbolized
each proper vertex by a blob.
 Diagrams involving ghost lines can be excluded for the same reasons as 
before.

Using Property 2 it is easy to verify that if the external momenta are slightly
off-shell, the momenta carried by the propagators in figg.
 (\ref{tree}),  (\ref{trunc}) also are, and the scalar products of all
momenta are of order $\omega$; $\omega$ is the infrared regulator
(with dimension of a mass) introduced in section 1. The exact propagators
for massless particles in the infrared limit have to behave as the naive
ones, i.e.\ they are of order $\omega^{-2}$.

Let  $E_{\gamma},E_y,E_g$ and $I_{\gamma},I_y,I_g$
denote respectively the number of external and internal photon,YM boson,
graviton lines coming out of one of the diagrams in fig. (\ref{trunc}).
Let $m^v_{\gamma},m_y^v,m^v_g$ denote the numbers of photons,YM bosons,
gravitons coming out from the $v^{th}$ proper vertex $\Gamma^v$ appearing in
the same diagram. Clearly,
\begin{eqnarray}
E_{\gamma} & = & \sum_v m^v_{\gamma}-2I_{\gamma}\nonumber \\
 E_y & = & \sum_v m_y^v-2I_y \nonumber \\
 E_g & = & \sum_vm^v_g-2I_g.
\label{exter1}
\end{eqnarray}
Moreover,
\begin{eqnarray}
N_p -I_p \ge \theta(E_p)
\label{exter2}
\end{eqnarray}
where $\theta(x):=\cases{0 \mbox{   if  $x=0$ }\cr  1 \mbox{   if  $x>0$ }\cr}$
and
$N_p$  denotes the number  of proper vertices in the diagram
where at least one particle $p$ ($p$ being a YM boson and/or a graviton) comes
out; this
unequality follow from the fact that $N_p=0$ if and only if $E_p=0=I_p$.

The results of the next section (Property 10) can be 
summarized  as follows, that
\begin{equation}
\Gamma^v= o(\omega^{m^v_{\gamma}+4\theta(m_y^v)-m^v_y+2\theta(m_g^v)
\delta_0^{m^v_y}})
\end{equation}
where in our notation $o(\omega^p)$ will denote an infinitesimal or an
infinite of {\it at least} order $p$ in $\omega$,
namely
$\lim\limits_{\omega\rightarrow 0}[o(\omega^p)\omega^{-p}]$ is zero or finite.
The overall $\omega$-dependence of the diagram contribution $D(\omega)$
will be the product
of the dependences of each vertex and each propagator:
\begin{equation}
D(\omega)=\left[\prod_v o(\omega^{m^v_{\gamma}+4\theta(m^v_y)-m^v_y+
2\theta(m^v_g)\delta_0^{m^v_y}}) \right] \omega^{-2(I_{\gamma}+I_y+I_g)}
\end{equation}
Using equations (\ref{exter1}), the latter becomes
\begin{equation}
D(\omega)=o(\omega^{E_{\gamma}-E_y+4(N_y-I_y)+2(N'_g-I_g)}),
\end{equation}
where $N'_g$  denotes the number  of proper vertices in the diagram where at
least one
graviton and no YM boson come out. To estimate $4(N_y-I_y)+2(N'_g-I_g)$ let
us
distinguish two cases. If $E_y=0$, then by colour conservation $m_y^v=0$ for
all vertices
in the diagram, implying $N_g'=N_g$; using  formulae (\ref{exter2}) for $p=y$
and $p=g$, we find
$4(N_y-I_y)+2(N'_g-I_g)\ge 4\theta(E_y)+2\theta(E_g)$. If $E_y>0$, noting that
$(N_y+N_g')=N_p$, $I_y+I_g=I_p$, where now $p$ denotes {\it either} $y$ {\it
or}
$g$, and using
formulae (\ref{exter2}) , we find
$4(N_y-I_y)+2(N'_g-I_g)\ge 2\theta(E_y)+2\theta(E_p)=4\theta(E_y)$.
Summing up,
$4(N_y-I_y)+2(N'_g-I_g)\ge 4\theta(E_y)+2\theta(E_g)\delta_0^{E_y}$
This expression  depends only on the
numbers of external bosons of the process, not on the particular diagram
we are considering, therefore we find the following

\medskip \noindent {\bf Property 7.} -- The amplitude $T$ of a decay process
with $E_{\gamma}$ external photons, $E_y$ external YM boson and $E_g$
gravitons satisfies the condition:
\begin{equation}
T=o(\omega^{E_{\gamma}-E_y+4\theta(E_y)+2\theta(E_g)\delta_0^{E_y}}).
\label{prop9}
\end{equation}
This formula is valid
at any loop order in all particles different from the gravitons
and at least at one loop order in the gravitons,
because the matter action with a background metric is multiplicatively
renormalizable
\cite{freed}, whereas at first order in the graviton loops pure QG is finite
on-shell.

Note that the RHS of formula (\ref{prop9}):  1) is independent of the number of
external gravitons,
provided $E_y>0$; 2) vanishes if $E_y=0$.

\section{Ward identities}

In QED the proper $n$-photon vertices 
$\Gamma_n^{\mu_1...\mu_n}(p^1,...,p^n)$
satisfy the Ward identity
\begin{equation}
  p_{\mu_1}\Gamma_n^{\mu_1\mu_2...\mu_n}(p^1,p^2,...,p^n)
  \varepsilon_{\mu_2}(p^2)...\varepsilon_{\mu_n}(p^n)=0,
\end{equation}
where $p^i$ is the momentum of the $i$-th photon and
$\varepsilon_{\mu_i}(p^i)$ the corresponding polarization vector;
this transversality condition amounts to the gauge invariance of any
physical process involving $n$ (incoming or outgoing) photons.

In this section we first derive the identity above and its analogues
for general Yang-Mills (YM) and Einstein (with $\Lambda=0$)
Quantum Gravity (QG) theories in the momentum configuration of decay
processes (compare with Property 2). Then we  use them and a continuity
argument
to show that the proper vertex for any decay process with fixed external
momenta
vanishes in QED and QG, whereas it is finite in YM. The Ward identities
are derived formally by using naive functional integration considerations
based only on the gauge invariance of the classical action (not on its
explicit form).
In the case of QED,YM, their validity extends to the true (i.e.
renormalized) theories at any order in the loops because renormalization
preserves Ward identities. In the case of QG, their validity is guaranteed
at any loop order in the matter fields and at least at one loop order in the
gravitons,
because the matter action with a background metric is multiplicatively
renormalizable
\cite{freed}, whereas at first order in the graviton loops pure QG is finite
on-shell.

We start by fixing the notation. Let ${\cal{S}}_{inv}(\phi)$ denote the
(local) action depending on the classical fields $\{ \phi_I \}$ and
$R^I_{\alpha}(\phi)$ corresponding (local) gauge generators:
\begin{equation}
  \delta_{\xi}{\cal S }_{inv}={ {\delta {\cal S}_{inv}}
  \over{\delta \phi_I}}\delta_{\xi}\phi^I=0,
\end{equation}
We employ a condensed notation in which a capital index $I$
is a collective index; it represents both discrete indices and a
continuous space-time variables $x$. A repeated index
implies summation over discrete indices and integration over $x$.
Explicitly, in the case of QED, YM, QG the fields $\phi_I$ include
\begin{equation}
  \phi_I:=\cases{
  A_{\mu}(x), \ \psi(x),\bar{\psi}(x) \ \
  \mbox{and/or $\varphi(x)$, $\bar{\varphi}(x)$ in QED;} \cr
  A^a_{\mu}(x), \ \ \mbox{+ possibly $\psi^i(x), \, \bar{\psi}^i(x)$
  and/or $\varphi^i(x), \, \bar{\varphi}^i(x)$ in YM;} \cr
  h_{\mu\nu}(x) \ \ \mbox{+ possibly any $\phi_I$ considered in the two
  previous cases in QG;} \cr}
\end{equation}
$x\in M^4$ denotes the point in Minkowski
spacetime, $A_{\mu}(x),A^a_{\mu}(x)$ the gauge potentials corresponding
respectively to a $U(1)$ and a semisimple group $G$,  $\psi(x),\bar{\psi}(x)$
(resp.\ $\varphi(x),\bar{\varphi}(x)$) spinors (complex scalars),
$\psi^i(x),\bar{\psi}^i(x)$ (resp.\ $\varphi^i(x),\bar{\varphi}^i(x)$)
spinors (complex scalars) making up a finite multiplet belonging to some
finite representation $Rep({\cal L}ie(G))$ (in the latter case $(T^a)^i_j$
will denote the matrix representation of the hermitean Lie algebra generators
corresponding to $A^a_{\mu}$), $h_{\mu\nu}(x)$ is the graviton field,
 $\eta_{\mu\nu}$ denotes the Minkowski metric tensor
(which plays the role of background metric) in cartesian coordinates,
and $g_{\mu\nu}(x)=\eta_{\mu\nu}+\kappa h_{\mu\nu}$ is the
the metric tensor.
The invariant actions ${\cal S }_{inv}$ read
\begin{equation}
  {\cal S }_{inv} = \cases{
  - \frac{1}{4} \int_{M^4} d^4x \, (F^{\mu\nu}F_{\mu\nu})
  +{\cal S}_{mat} \qquad \qquad \mbox{in QED;} \cr
  - \frac{1}{4} \int_{M^4} d^4x \, (F^{a~\mu\nu}F^a_{\mu\nu})
  +{\cal S}_{mat} \qquad \qquad \mbox{in YM;} \cr
  \int_{M ^4} d^4x \, g^{\frac{1}{2}}(\lambda -\frac{1}{16\pi G}R)
  +{\cal S}_{mat} \qquad \qquad \mbox{in QG},\cr}
\label{act1}
\end{equation}
where $F_{\mu\nu},F^a_{\mu\nu}$ is the field strenght in QED,YM 
respectively,
$R$ is the Ricci scalar of the metric $g_{\mu\nu}$, $g:=-det[g_{\mu\nu}]$,
$f^{abc}$ are the structure constants of ${\cal L}ie(G)$ and $e$ the coupling
constant. ${\cal S}_{mat}$ is the action of the matter  minimally
coupled to the gauge potential
\footnote{Strictly speaking, in the case of QG an action  ${\cal S}_{mat}$
containing a spinor contribution requires the introduction of
vierbeins as dynamical variables instead of the metric. However, the
considerations of this section hold also in that case, since they are based
on the gauge tranformations (\ref{gaugetr3}) of the metric, which
can be obtained from the gauge transformations of the vierbeins.}.

$A_{\mu},A^a_{\mu},h_{\mu\nu}$ are respectively the gauge potentials for
QED, YM, QG, with  gauge transformations
\begin{eqnarray}
  \delta_\xi A_{\mu} & = & \partial_{\mu}\xi \qquad \mbox{in QED;}
  \label{gaugetr1} \\
  \delta_\xi A^a_{\mu} & = & (D_{\mu}\xi)^a:=\partial_{\mu}\xi^a
  + e f^{abc}A^b_{\mu}\xi^c \qquad \mbox{in YM;}
  \label{gaugetr2} \\
  \delta_\xi g_{\mu\nu} & = & g_{\nu\rho}\partial_{\mu}\xi^{\rho}
  + g_{\mu\rho}\partial_{\nu}\xi^{\rho}+ \xi^{\rho}
  \partial_{\rho}g_{\mu\nu},    \nonumber\\
\delta_\xi A^a_{\mu} & = & A^a_{\rho}\partial_{\mu}\xi^{\rho}+\xi^{\rho}
\partial_{\rho}A^a_{\mu}
\qquad\qquad\qquad \mbox{in QG} .
  \label{gaugetr3}
\end{eqnarray}
We omit for the sake of brevity the well-known gauge transformations of the
other fields.

The quantization of the theory (in a perturbative setting) is performed in the
BRST formalism \cite{brst,bv}: the set of fields
$\{ \phi_I\}$ is enlarged to a set $\{ \Phi_A \}$ by the introduction of
ghosts, antighosts and Stueckelberg fields, and we associate to the action
${\cal S}_{inv}$  a gauge-fixed action $S_{\Psi}$ depending on the
gauge-fixing functional $\Psi$. Index $A$, like $I$,  represents both
discrete indices and the continuous space-time variables $x$.
Let $S_{GF}:=S_{\Psi}(\Phi)-{\cal S }_{inv}(\phi)$; in QED and YM, $S_{GF}$ can
be constructed as $S_{GF}=s\Psi$, where $s$ denotes the BRST transformation
associated to the gauge transformations (\ref{gaugetr1}) - (\ref{gaugetr3}).

The generating functional $Z(J)$ (depending on the external sources $J$)
for the Green functions of the theory is defined by
\begin{equation}
  Z(J):= \int {\cal{D}}\Phi e^{\frac{i}{\hbar}
  [S_{\Psi}(\Phi)+J^A \Phi_A]},
\label{zeta}
\end{equation}
where ${\cal{D}}\Phi$ is a gauge invariant functional measure,  $J^A$
transforms
under diffeomorphisms as the appropriate tensor density.

By performing a gauge \footnote{Alternatively, one could perform a BRST
transformation; the resulting Ward identities would be the same.}
transformation $\phi\rightarrow \phi+\delta_{\xi}\phi$
of the dummy integration variables $\phi$ in the RHS of eq.\ (\ref{zeta})
the integral $Z(J)$ remains the same (the Jacobian is 1),
implying the Ward identities
\begin{equation}
  0=\delta_{\xi}Z(J)=\frac{i}{\hbar}
  \int {\cal{D}}\Phi [J^A \, \delta_{\xi}\Phi_A+ \delta_{\xi} S_{GF}]
  e^{\frac{i}{\hbar} [S_{\Psi}(\Phi)+J^A\Phi_A ]},
\end{equation}
or, in terms of the generating functional $W(J):=\frac{\hbar}{i} ln[Z(J)]$
of the connected Green functions,
\begin{equation}
  0=\left. \left[J^A \, \delta_{\xi}\Phi_A + \delta_{\xi} S_{GF}
  \right] \right|_{\Phi_A\rightarrow
  \frac{\delta}{\delta J^A}} \, W(J)+ disconnected~~terms.
\end{equation}
The disconnected terms are absent  when evaluating the
Green function on any decay process, since in this case only {\it one}
initial particle is present. Therefore, as far as we are concerned,
\begin{equation}
  0=\left. \left[J^A \, \delta_{\xi}\Phi_A + \delta_{\xi} S_{GF}
  \right] \right|_{\Phi_A\rightarrow
  \frac{\delta}{\delta J^A}} \, W(J).
\label{waw}
\end{equation}
In order to obtain the Ward identities for the proper vertex functions we
introduce the usual Legendre transform $\Gamma(\tilde{\Phi}):=
[W(J)-J^A\Phi_A]|_{J=J(\tilde{\Phi})}$,
where the function $J=J(\tilde{\Phi})$ is obtained by inverting the relations
$\tilde{\Phi}_A=\frac{\delta W}{\delta J^A}$; the new independent
variables are the ``classical fields'' $\tilde{\Phi}$. Consequently
$J^A(\tilde{\Phi})= -\frac{\delta \Gamma}{\delta \tilde{\Phi}_A}$.

{}From identity (\ref{waw})
we draw the following Ward identities for the generating functional of
proper vertices $\Gamma$
\begin{equation}
  0=\left[\frac{\delta\Gamma}{\delta\tilde{\Phi}_A} \cdot \delta_{\xi}
  \tilde{\Phi}_A+ \delta_{\xi} S_{GF}(\tilde{\Phi})\right].
\label{wag}
\end{equation}

Actually, we are interested in the Ward identities for the proper vertices
having only physical gauge bosons as external (incoming or outcoming)
particles. The physicality condition is best imposed in momentum space. The
proper vertex $\Gamma_n^{12...n}(x^1,x^2,...,x^n)$ with $n$ external gauge
bosons $b_i(x^i)$ (in configuration space) is obtained from $\Gamma$
through differentiation,
\begin{equation}
  \Gamma_n^{12...n}(x^1,x^2,...,x^n) = \left.
  \frac{\delta^n\Gamma}{\delta b_1(x^1)...\delta b_n(x^n)}
  \right|_{\tilde{\Phi}=0},
\end{equation}
where we have introduced the short-hand notation
\begin{equation}
  i\rightarrow\cases{\mu_i \cr (\mu_i,a_i) \cr \mu_i\nu_i \\\ or\\\
(\mu_i,a_i)\cr} \qquad
  b_i\rightarrow\cases{\tilde A_{\mu_i} \ \ \ \mbox{in QED} \cr
  \tilde A^a_{\mu_i} \ \ \ \mbox{in YM} \cr \tilde h_{\mu_i\nu_i}\\\ or\\\
A^a_{\mu_i}
  \ \ \ \mbox{in QG;}\cr} \qquad \qquad i=1,2,...,n.
\end{equation}
The RHS has automatically the required boson symmetry in the identical
particles,
e.g. if all the $b_i$'s are the same type of fields
\begin{equation}
  \Gamma_n^{i_1i_2...i_n}(x^{i_1},x^{i_2},...,x^{i_n})=
  \Gamma_n^{12...n}(x^1,x^2,...,x^n),
\label{bosym}
\end{equation}
where $(i_1,i_2,...i_n)$ is a permutation of $(1,2,...,n)$. On account of
the translation invariance $\Gamma_n^{1...n}(x^1,...,x^n)=
\Gamma_n^{1...n}(x^1+a,...,x^n+a)$, its multiple Fourier transform can be
written as $\Gamma_n^{1...n}(p^1,...,p^n)\delta^4(\sum\limits_{i=1}^n p^i)$;
it contains  a Dirac-$\delta$ implementing the total momentum conservation.
Here and below our conventions for the Fourier transform will be
$f(p):=\int \frac{d^4x}{(2\pi)^4} \, e^{-ip\cdot x}f(x)$, \ $f(x)=\int
d^4p \, e^{ip\cdot x}f(p)$. As a consequence of the general relation
\begin{equation}
  \int \frac{d^4x}{(2\pi)^4} \, e^{-ip\cdot x}\frac{\delta
  F}{\delta\phi(x)}=(2\pi)^{-4}\frac{\delta F}{\delta\phi(-p)}
\end{equation}
one finds
\begin{equation}
  \delta^4 \left( \sum\limits_{i=1}^n p^i \right)
  \Gamma_n^{12...n}(p^1,p^2,...,p^n)=
  (2\pi)^{-4n} \left. \frac{\delta^n\Gamma}{\delta b_1(-p^1)...
  \delta b_n(-p^n)}
  \right|_{\tilde{\Phi}=0}.
\end{equation}

Differentiating relation (\ref{wag}) with respect to
$b_1(-p^1),...,b_n(-p^n)$ and setting thereafter $\tilde{\Phi}=0$, we obtain
\begin{eqnarray}
  0 & = &\int d^4q \left[(2\pi)^4  \delta^4\left(q+\sum\limits_{j=1}^n
p^{j}\right)
 \Gamma_{n+1}^{01...n}(q,p^1,...,p^n)\delta_{\xi}b_0(q) \right.\nonumber \\
& &+ \sum\limits_{h=1}^n
  \delta^4\left(q+\sum\limits_{j=1,~j\neq h}^n p^j\right)
   \Gamma_n^{01...,h\!-\!1,h\!+\!1,...n}(q,p^1,...,p^{h-1},p^{h+1},...,p^n)
  \frac{\delta (\delta_{\xi}b_0(q))}
  {\delta b_h(-p^h)}
   \nonumber \\
  & & \ \ + \left. \left. \frac{\delta^n \delta_{\xi}S_{GF}(\tilde{\Phi})}
  {\delta b_1(-p^1)...\delta b_n(-p^n)} \right]
  \right|_{\tilde{\Phi}=0}.
\label{wab}
\end{eqnarray}

In fact,
 only the terms with $\tilde{\Phi}_A=b$ in
  the first term in eq.\ (\ref{wag}) contribute to eq (\ref{wab}),
since when $\tilde{\Phi}_A\neq b$
  then $\frac{\delta^m(\delta_{\xi}\tilde{\Phi}_A )}{\delta b_1...\delta b_m}
  |_{\tilde{\Phi}=0}=0$
  (indeed, for any $\Phi_A$ $\delta_\xi\Phi_A$ is of degree
  $\ge 1$ in $\Phi_A$).

To get identities involving proper vertices with {\it physical} external
bosons we will have to  contract
their Lorentz indices with the ones of  transverse polarization tensors/vectors
(we will choose them with well-defined helicity) $e^1(p^1)...e^n(p^n)$, where
\begin{equation}
  e(p)=e^{\pm}(p):=\cases{ \varepsilon^{\pm}_{\mu}(p) \qquad
  \mbox{when $b=\tilde A_{\mu},\tilde A^a_{\mu}$} \cr
  (\varepsilon^{\pm}_{\mu}(p)\varepsilon^{\pm}_{\nu}(p)) \qquad
  \mbox{when $b=\tilde h_{\mu\nu}$},\cr} \qquad \qquad
  \mbox{with $\varepsilon^{\pm}_{\mu}(p)p^{\mu}=0$}.
\label{trans}
\end{equation}

Now it is easy to realize that in all cases the following property holds:
\begin{equation}
  \left. \frac{\delta^n \delta_{\xi}S_{GF}(\tilde{\Phi})}
  {\delta b_1(p^1)...\delta b_n(p^n)}
  \right|_{\tilde{\Phi}=0}e_1^1(-p^1)...
  e_n^n(-p^n)=0;
\label{appen}
\end{equation}
where contraction of the Lorentz indices hidden in  the symbols $1,...,n$ and
$e^1,...,e^n$ is understood.
In fact, the terms of non-zero degree in the ghosts contained in
$\delta_{\xi}S_{GF}$ vanish after setting $\tilde{\Phi}'=0$; the other terms
depend on the longitudinal modes of the bosons, and vanish after
contraction with the polarization vectors/tensors. We prove  explicitly
this statement in the appendix, for the Feynman (harmonic) gauge fixings.

Introducing the notation
\begin{equation}
  \Gamma_n^{1...e^i...n}:=\Gamma_n^{1...i...n}\cdot e^i,
\end{equation}
where again contraction of the Lorentz indices hidden in the symbols $i$ and
$e^i$ is understood, the Ward identities (\ref{wab2}) will therefore reduce to
\begin{eqnarray}
  0 & = &\int d^4q \left[(2\pi)^4  \delta^4\left(q+\sum\limits_{j=1}^l
p^{j}\right)
 \Gamma_{n+1}^{0e^1...e^n}(q,p^1,...,p^n)\delta_{\xi}b_0(q) \right.
\nonumber \\ & &+ \sum\limits_{h=1}^n
  \left. \left.\delta^4\left(q+\sum\limits_{j=1,~j\neq h}^n p^j\right)
   \Gamma_n^{0e^1...,e^{h\!-\!1},e^{h\!+\!1},...e^n}(q,p^1,...,p^{h-1},
p^{h+1},...,p^n)  \frac{\delta (\delta_{\xi}b_0(q))}
  {\delta b_h(-p^h)}e^h \right]
  \right|_{\tilde{\Phi}=0}.
\label{wab2}
\end{eqnarray}

The identity above is one essential ingredient that we need in order to prove
the main
property of this section.  In order to formulate this property, we need now a
notion of ``vicinity''
of a ``decay configuration'' parametrized by one regularization parameter
$\omega$.
Therefore, we introduce some useful definitions.

A configuration $\omega$-converging  to  the decay configuration
$( \hat{k}^i, \hat{\varepsilon}^i)_{i=0,...,n}$ ($\omega\ge 0$)
is a  one-parameter family
$(k^i(\omega), \varepsilon^i(\omega))_{i=0,...,n}$
 such that $\varepsilon^i(\omega)\cdot k^i(\omega)=0$,
$k^i(\omega)- \hat k^i=o(\omega)$,
$\varepsilon^i(\omega)- \hat \varepsilon^i=o(\omega)$,
$k^i\cdot k^{i'}=o(\omega^2)$
$\forall i,i'=0,1,...,n$.
Examples of these families will be given in formulae (\ref{fam1}),
(\ref{fam2}).

 It is easy to show that in the mentioned hypotheses  the 3-momenta
are in general no more collinear, but form angles $\siml \omega$; consequently,
\begin{equation}
\varepsilon^i(k^i)\cdot\varepsilon^j(k^j)=
\cases{either~~o(1) \cr or~~o(\omega) \cr} \qquad\qquad
\varepsilon^i(k^i)\cdot k^j=o(\omega). \qquad\qquad
\label{approx}
\end{equation}

We are now able to prove the following fundamental property of the vertices,
which is the main result of this Section and adds to the kinematical
properties of Section 1:

\medskip \noindent {\bf Property 10.} --
On any configuration $(k^i(\omega), \varepsilon^i(\omega))_{i=0,...,n}$
$\omega$-converging
to the decay configuration $( \hat{k}^i, \hat{\varepsilon}^i)_{i=0,...,n}$
\begin{eqnarray}
  & & \Gamma_{n+1}^{e^0....e^n}(k^0,...,k^n)= o(\omega^{n+1}) \qquad
\mbox{in QED;} \label{main1} \\
  & & \Gamma_{n+1}^{e^0a_0...e^na_n}(k^0,...,k^n)= o(\omega^{4-n-1})
\qquad \mbox{in YM;}
\label{main2}\\
  & & \Gamma_{n+1}^{e^0...e^n}  (k^0,...,k^n)=o(\omega^{m_{\gamma}+
\theta(m_y)(4-m_y)+2\theta(m_g)\delta_0^{m_y}})
\qquad \mbox{in QG.}
\label{main3}
\end{eqnarray}
where in the third equation $m_{\gamma},m_y,m_g$ denote the number of
external photons,
YM bosons and gravitons respectively ($m_{\gamma}+m_y+m_g=n+1$),
and $\theta(x):=\cases{0\mbox{  if  $x=0$} \cr 1\mbox{  if  $x>0$} \cr}$.

\vskip1truecm

\noindent
{\it Proof}.

The claim is evidently true when $n=0$. In fact,
$\Gamma_1^{\mu_0}\propto (k^0)^{\mu_0}$
in QED, YM, but this vanishes since momentum conservation
imposes the condition $k^0=0$; in QG still it could be
$\Gamma_1^{\mu_0\nu_0}=const \times \eta^{\mu_0\nu_0}$, but this vanishes
after
contraction with $e^{\mu_0\nu_0}$ (which is a traceless tensor).

The rest of the proof is by induction and divided in three parts. Let us assume
that the
claim is true when $n=m-1$. We will prove that it is true when $n=m$.
For the sake of simplicity, we explicitly prove the claim (\ref{main3}), which
is the
most general possible,
in the simpler case $m_{\gamma}=0=m_y$,
\begin{equation}
\Gamma_{n+1}^{e^0...e^n}  (k^0,...,k^n)= o(\omega^2)  \qquad \mbox{in QG};
\label{main3s}
\end{equation}
at the end of this section we will briefly sketch how the proof goes in the
general
case.

{\bf Part 1} Here we prove the equations
\begin{eqnarray}
  & & \Gamma_{n+1}^{e^0...e^{i-1},\mu_i,e^{i+1}...e^n}(k^0,...,k^n)
  k_{\mu_i}^i = 0 \qquad \mbox{in QED;} \label{wai1} \\
  & &\Gamma_{n+1}^{e^0a_0...e^{i-1}a_{i-1},\mu_ia_i,e^{i+1}a_{i+1}...e^na_n}
  (k^0,...,k^n)k_{\mu_i}^i = o(\omega^{4-n}) \qquad \mbox{in YM;} \label{wai2}
\\
  & & \Gamma_{n+1}^{e^0...e^{i-1},\mu_i\nu_i,e^{i+1}...e^n}
  (k^0,...,k^n)k_{\mu_i}^i = o(\omega^2)  \qquad \mbox{in QG.}
\label{wai3}
\end{eqnarray}
We drop in the sequel the tilde and write $A_{\mu},A^a_{\mu},
g_{\mu\nu}$ instead of $\tilde A_{\mu},\tilde A^a_{\mu},\tilde g_{\mu\nu}$.
We treat separately the cases of QED, YM and QG.

\medskip \noindent --  QED.
{}From $\delta_{\xi}A_{\mu}(p)=ip_{\mu}\xi(p)$ (eq.\ (\ref{gaugetr1})),
and eq.\ (\ref{wab2}), from differentiating w.r.t.\ $q$ it immediately follows
\begin{equation}
  p^0_{\mu_0}\Gamma_{n+1}^{\mu_0e^1...e^n}(p^0,p^1,...,p^n)=0
\label{waiqed}
\end{equation}
(we have factored out $\delta^4(\sum\limits_{i=0}^n p^i)$),
whence formula (\ref{wai1}) follows at once (using boson symmetry), if we
choose $p^i$ so that the sets $\{p^0,...,p^n\}$, $\{k^0,...,k^n\}$ coincide.
Actually we can derive directly from eq. (\ref{wab}) the stronger property
\begin{equation}
  k^i_{\mu_i}\Gamma_{n+1}^{\mu_0...\mu_i...\mu_n}(k^0,...,k^i,...,k^n)=0,
\qquad\qquad n\ge 2
\label{waiqedbis}
\end{equation}

\medskip \noindent --  YM.  From
\begin{equation}
  \delta_{\xi}A^a_{\mu}(p)=ip_{\mu}\xi^a(p)
  +e f^{abc}\int d^4q \, A_{\mu}^b(p-q)\xi^c(q)
\end{equation}
(eq.\ (\ref{gaugetr2}) in momentum space), and from differentiating
formula (\ref{wab2}) (with $n=m$) w.r.t.\ $\xi(p^0)$, it immediately follows
\begin{eqnarray}
& & ip^0_{\mu_0}\Gamma_{m+1}^{\mu_0a_0,e^1a_1,...,e^ma_m}(p^0,p^1,...,p^m)
  + \nonumber \\
& & \ \ +\sum\limits_{l=1}^m ef^{b_la_la_0}
  \Gamma_m^{e^1a_1,...e^{l-1}a_{l-1},e^lb_l,e^{l+1}a_{l+1},...,
  e^ma_m}(p^1,...,p^{l-1},p^l+p^0,p^{l+1},...,p^m)=0
\end{eqnarray}
(again, we have factored out $\delta^4(\sum\limits_{i=0}^m p^i)$). This
formula holds for any  configuration
$\sum\limits_{i=0}^m p^i=0$, $e^i(p^i)\cdot p^i=0$. On a configuration
$\omega$-converging
to the decay configuration  we deduce from the induction hypothesis that the
second term is $o(\omega^{4-m})$.

\medskip \noindent --  QG. The gauge transformation (\ref{gaugetr3}) in
momentum space reads
\begin{equation}
  \delta_{\xi}g_{\mu\nu}(p) =  i \int d^4r  \Bigl\{g_{\rho\nu}(p-r)r_{\mu}
  \xi^{\rho}(r)+g_{\rho\mu}(p-r)r_{\nu} \xi^{\rho}(r)+\xi^{\rho}(p-r)r_{\rho}
g_{\mu\nu}(r)
  \Bigr\} ,
\label{deltag}
\end{equation}
implying
\begin{equation}
  \left. \delta_{\xi}g_{\mu\nu}(p)
  \right|_{g_{\mu\nu}(p)=\eta_{\mu\nu}\delta^4(p)}=
  i\{p_{\mu}\xi^{\rho}(p)\eta_{\rho\nu}+p_{\nu} \xi^{\rho}(p)\eta_{\rho\mu}\}.
\end{equation}
Moreover, we note that
\begin{equation}
  \frac{\delta g_{\alpha\beta}(p)}{\delta
  g_{\mu\nu}(-q)}=\delta^4(p+q)\left[\delta_{\alpha}^{\mu}
  \delta_{\beta}^{\nu}+\delta_{\alpha}^{\nu}\delta_{\beta}^{\mu}\right].
\label{deriv}
\end{equation}

After differentiation
w.r.t.\ $\xi^{\nu_0}(p^0)$, Eq.\ (\ref{wab2}) with $n=m$ reads:
\begin{eqnarray}
  0 & = &  \Gamma_{m+1}^{\mu\nu,e^1,...,e^m}(p^0,p^1,...,p^m)
  2(p^0)_{\mu}\eta_{\nu\nu_0} \nonumber \\
  & & +\sum\limits_{h=1}^m\left[
  \Gamma_m^{e^1...e^{h-1},\mu\nu,e^{h+1}...e^n}(...,p^{h-1},p^0+p^h,
p^{h+1},...)4(p^0)_{\mu}
(\varepsilon^h)_{\nu}(\varepsilon^h)_{\nu_0}
  \right. \nonumber \\
  & & \qquad \left. + \left. (p^h)_{\nu_0}
 \Gamma_m^{e^1...e^n}(...,p^{h-1},p^0+p^h,p^{h+1},...)\right]
    \right|_{g_{\alpha\beta}(p)=\eta_{\alpha\beta}\delta^4(p)}.
\label{long}
\end{eqnarray}
(once again, we have factored out $\delta^4(\sum\limits_{i=0}^m p^i)$). This
formula holds for any  configuration
$\sum\limits_{i=0}^m p^i=0$, $e^i(p^i)\cdot p^i=0$. On a configuration
$\omega$-converging
to the decay configuration  we deduce from the induction hypotheses
 (\ref{wai3}), (\ref{main3s}) that the
second, third terms are $o(\omega^2)$, which proves eq. (\ref{wai3}) for $n=m$.

\vskip1truecm

{\bf Part 2:}
We prove the factorization formulae
\begin{eqnarray}
  & & \Gamma_{n+1}^{e^0....e^n}(k^0,...,k^n)= \sum\limits_P
A_{i_0i_1...i_n} E^{i_0i_1}...E^{i_{n-1}i_n}\qquad \mbox{in QED, (n+1) even:}
\label{fact1} \\
  & & \Gamma_{n+1}^{e^0a_0...e^na_n}(k^0,...,k^n)=  \sum\limits_P
A^{a_0...a_n}_{i_0i_1...i_n} E^{i_0i_1}...E^{i_{n-1}i_n}+o(\omega^{3-n}) \qquad
\mbox{in YM, (n+1) even;}
\label{fact2}\\
  & & \Gamma_{n+1}^{e^0...e^n}  (k^0,...,k^n)= \sum\limits_P
A_{i_0i_1...i_{2n}i_{2n+1}} E^{i_0i_1}...E^{i_{2n}i_{2n+1}}+ o(\omega)  \qquad
\mbox{in QG.}
\label{fact3}\
\end{eqnarray}
and
\begin{eqnarray}
  & & \Gamma_{n+1}^{e^0....e^n}(k^0,...,k^n)=
\sum\limits_P\sum\limits_{j_0=0}^n
A^{j_0}_{i_0i_1...i_n}
(k^{j_0}\cdot\varepsilon^{i_0})E^{i_1i_2}...E^{i_{n-1}i_n}
\qquad
\mbox{in QED, (n+1) odd;}
\label{factb1} \\
  & & \Gamma_{n+1}^{e^0a_0...e^na_n}(k^0,...,k^n)=  \sum\limits_P
\sum\limits_{j_0=0}^n
A^{j_0;a_0...a_n}_{i_0i_1...i_n} (k^{j_0}\cdot\varepsilon^{i_0})
E^{i_1i_2}...E^{i_{n-1}i_n}+o(\omega^{3-n}) \qquad \mbox{in YM, (n+1) odd;}
\nonumber \\
& & 
\label{factb2}
\end{eqnarray}
where:

1) $\sum\limits_P$ means the sum  over all the permutations $P$
($P\equiv(i_0,i_1,...,i_n)$ is a permutation
of $(0,1,...,n)$ in QED and YM,
whereas $P\equiv(i_0,i_1,...,i_{2n+1})$ is a permutation of $(0,1,...,2n+1)$ in
QG);

2) the
$A$'s are scalar functions depending on the scalar products $k^i\cdot k^j$
(and, in the Y.M. case, on $2m$ Lie algebra indices $a_i$);

3) we have introduced the shorthand notation
\begin{equation}
E^{ij}:= \left(\varepsilon^i\cdot\varepsilon^ j~k^i\cdot k^j-
\varepsilon^i\cdot
 k^j~\varepsilon^j\cdot  k^i    \right).
\end{equation}
In the RHS of eq. (\ref{fact3}) it is tacitly understood that
$\varepsilon^{2s+1}\equiv\varepsilon^{2s}$,
$k^{2s+1}=k^{2s}$, $s=0,..., 2n$.

We prove explicitly the first three (the proof of formulae (\ref{factb1}),
(\ref{factb2}), is completely
analogous): let $n+1=2m$.
We look for the most general
$\Gamma_{n+1}^{\mu_1....\mu_{n+1}}(k^1,...,k^{n+1})$  satisfying:

1)  the constraint
\begin{equation}
  \Gamma_{n+1}^{\varepsilon^0...\varepsilon^{i-1},\mu_i,\varepsilon^{i+1}...
\varepsilon^n}(k^0,...,k^n)
  k_{\mu_i}^i = o(\omega^d)  \label{wai?}
\end{equation}
in any configuration $(k^i(\omega), \varepsilon^i(\omega))_{i=0,...,n}$
$\omega$-converging
to the decay configuration $( \hat{k}^i, \hat{\varepsilon}^i)_{i=0,...,n}$;

2) symmetry under any replacement $(\mu_i,k^i)\leftrightarrow(\mu_l,k^l)$,
$i,l=0,...,n$.

If we set $o(\omega^d)\equiv 0$ this amounts to solving eq. (\ref{wai1})
equipped with boson symmetry for the $(n\!+\!1)$-photons vertex function
of  Q.E.D.; if we set $d=3-n$, this amounts to solving eq. (\ref{wai2})
equipped
with boson symmetry
for the $(n\!+\!1)$-gluons vertex function of Y.M., provided we  understand an
implicit dependence
of $\Gamma_{n+1}$ on the
Lie algebra indices $a_i$ and remind that the latter have to be permuted along
with the indices
$\mu_i$ and the momenta $k^i$ when boson symmetry is imposed; if we
choose $n\!+\!1=4r$, $d=2$,
and add the additional symmetry conditions
$k^{2i+1}= k^{2i}$, $\varepsilon^{2i+1}= \varepsilon^{2i}$ ($i=0,...,2r-1$),
this will amount to solving eq. (\ref{wai3}) equipped with boson symmetry
for the $2r$-gravitons vertex function of Q.G. In this way,
we can formally deal with eq.'s (\ref{wai1}), (\ref{wai2}),  (\ref{wai3})
simultaneously, by just dealing
with one.

The  dependence of $\Gamma_{n+1}^{\mu_0....\mu_n}(k^0,...,k^n)$ on
Lorentz indices can only occur
through the metric tensors $\eta^{\mu_i\nu_j}$ and the 4-vectors $k^{\mu_l}$.
Compactly, the most general
dependence can be written in the following way
\begin{equation}
\Gamma_{n+1}^{\mu_0....\mu_n}=\sum B^0\underbrace{\eta...\eta}_{m~times}
+\sum B^1kk\underbrace{\eta...\eta}_{(m\!-\!1)~times}+...+\sum B^m
\underbrace{k...k}_{2m~times},
\label{stru1}
\end{equation}
where the $B$'s denote Lorentz scalar functions. For our purposes, it
will be more convenient to expand $\Gamma_{n+1}$ in terms of the 4-vectors
$k^{\mu_l}$ and of the
tensors
$E^{\mu_i\mu_j}(k^i,k^j):=\eta^{\mu_i\mu_j}k^i\cdot k^j-(k^i)^{\mu_j}
(k^j)^{\mu_i})$,
which satisfy the relation
\begin{equation}
(k^i)_{\mu_i}E^{\mu_i\mu_j}=0=(k^j)_{\mu_j}E^{\mu_i\mu_j}
\label{transv}
\end{equation}
The general expansion (\ref{stru1}) can be replaced by
\begin{equation}
\Gamma_{n+1}^{\mu_0....\mu_n}(k^0,...,k^n)=\sum\limits_P\sum\limits_{l=0}^m
\sum\limits_{j_0,...j_{2l-1}=0}^n
 A^{l;~j_0...j_{2l-1}}_{i_0...i_n}(k^{j_0})^{\mu_{i_0}}...
(k^{j_{2l-1}})^{\mu_{i_{2l-1}}}
E^{\mu_{i_{2l}}\mu_{i_{2l+1}}}....E^{\mu_{i_{n-1}}\mu_{i_n}}
\label{stru2}
\end{equation}
where $\sum\limits_P$ means the sum  over all the permutations
$P\equiv(i_0,i_1,...,i_n)$ of $(0,1,...,n)$ and
$A^{l;~j_1...j_{2l}}_{i_0...i_n}$ are scalar functions depending on the
scalar products $k^i\cdot k^j$
(and, in the Y.M. case, on $2m$ Lie algebra indices $a_i$).

We have introduced a quite redundant set of scalars
$\{ A^{l;~j_1...j_{2l}}_{i_0...i_n}\}$ to make
formula (\ref{stru2}) more compact.  The set is redundant in the sense that
$A^{l;~j_0...j_{2l-1}}_{i_0...i_n}$ and
$A^{l;~\hat  j_0...\hat j_{2l-1}}_{\hat i_0...\hat i_n}$
will both contribute to the same term
$(k^{j_0})^{\mu_{i_0}}...(k^{j_{2l-1}})^{\mu_{i_{2l-1}}}
E^{\mu_{i_{2l}}\mu_{i_{2l+1}}}
...E^{\mu_{i_{n-1}}\mu_{i_n}}$ in the expansion  (\ref{stru2}), whenever

1)  there exists a permutation $P_{2l}$ of $2l$ objects such that
$(\hat i_0,\hat i_1,...,\hat i_{2l-1})=P_{2l}(i_0,i_1,...,i_{2l-1})$,
$(\hat j_0,\hat j_1,...,\hat j_{2l-1})=P_{2l}(j_0,j_1,...,j_{2l-1})$;

2) $(\hat i_{2l},\hat i_{2l+1},...,\hat
i_n)=P_{n+1-2l}(i_{2l},i_{2l+1},...,i_n)$, where
$P_{2m-2l}$ is a permutation  of $n+1-2l=2m-2l$ objects which is the product:
2.a) of
transpositions between the $(2s)^{th}$ and the $(2s+1)^{th}$ object
($s=1,...,m-l$); 2.b)
of transpositions between different pairs $(2s,2s+1)$, $(2r,2r+1)$,
$r,s=1,...,m-l$.

We are free to set $A^{l;~j_0...j_{2l-1}}_{i_0...i_n}=A^{l;~\hat  j_0...
\hat j_{2l-1}}_{\hat i_0...\hat i_n}$
in these cases.

Finally, boson symmetry (\ref{bosym}) implies that the scalars $A^l$
satisfy the relations
\begin{equation}
A^{l;~\tilde j_0...\tilde j_{2l-1}}_{...j...i...}(k^i\leftrightarrow k^j)=
A^{l;~j_0...j_{2l-1}}_{...i...j...}
\qquad\qquad \tilde h:= \cases{j~~~\mbox{if}~~h=i \cr i~~~\mbox{if}~~h=j \cr
h~~~
\mbox{if}~h\neq i,j\cr}
\end{equation}
for any pair of indices $i,j$.

Plugging the general expansion (\ref{stru2}) into Eq.  (\ref{wai?})  and using
relation (\ref{transv}) we find
\begin{eqnarray}
o(\omega^d) & = & \sum\limits_{P'}\sum\limits_{l=1}^m
\sum\limits_{j_0,...j_{2l-1}=0}^n
 \left[A^{l;~j_1...j_{2l}}_{i i_1...i_n}+A^{l;~j_1...j_{2l}}_{i_1 i...i_n}+...+
A^{l;~j_0...j_{2l-1}}
_{i_1...i_{2l-1}ii_{2l}...i_n}\right]
\times \nonumber \\
&~ &(k^i\cdot k^{j_0})~\varepsilon^{i_1}\cdot k^{j_1}...\varepsilon^{i_{2l-1}}
\cdot k^{j_{2l-1}}
E^{i_{2l}i_{2l+1}} ...E^{i_{n-1}i_n},
\label{wain}
\end{eqnarray}
where $\sum\limits_{P'}$ means the sum  over all the permutations
$P'\equiv(i_1,...,i_n)$ of
$(0,1,..., i-1,i+1,...,n)$,
 whereas
\begin{eqnarray}
 \Gamma_{n+1}^{\varepsilon^0...\varepsilon^n}(k^0,...,k^n)
& = &\sum\limits_{P'}\sum\limits_{l=0}^m\sum\limits_{j_0,...j_{2l-1}=0}^n
 \left[A^{l;~j_0...j_{2l-1}}_{i i_1...i_n}+A^{l;~j_0...j_{2l-1}}_{i_1
i...i_n}+...+
A^{l;~j_0...j_{2l-1}}_{i_1...i_{2l-1}
ii_{2l}...i_n}\right]
\times \nonumber \\
& ~ & (\varepsilon^{i}\cdot k^{j_0})
{}~\varepsilon^{i_1}\cdot k^{j_1}...\varepsilon^{i_{2l-1}}\cdot k^{j_{2l-1}}
E^{i_{2l}i_{2l+1}} ...E^{i_{n-1}i_n}.
\label{expr}
\end{eqnarray}
Note that the term $l=0$ has completely disappeared from the sum in eq.
(\ref{wain}), due to
eq. (\ref{transv}).

Let us fix the $xyz$ axes so that $k^0=(k^0_0,0,0,k^0_0)$
[according to property 2 this implies $k^j=\lambda^j(k^0_0,0,0,k^0_0)$,
$j=1,2,...n$];
we can always assume that the polarization vectors
$\hat{\varepsilon}^i$ are real and have the form
$\hat{\varepsilon}^i=(0,cos\theta^i,sen\theta^i,0)$.
We now start exploiting the available freedom in the choice (1) of  the angles
$\theta^i$ characterizing
the polarization vectors
$\hat{\varepsilon}^i$; (2)  of the  configuration $(k^i, \varepsilon^i)$
$\omega$-converging
to  $( \hat{k}^i, \hat{\varepsilon}^i)_{i=0,...,n}$.
A family of possible choices of the latter is
\begin{eqnarray}
& & k^i\equiv \hat{k}^i+\omega b^i\hat{\varepsilon'}^i \qquad\qquad
\hat{\varepsilon'}^i:=(0,-sen\theta^i,cos\theta^i,0)
\qquad\qquad i=0,1,...,n,\nonumber \\
& & \varepsilon^i\equiv\hat{\varepsilon}^i;
\label{fam1}
\end{eqnarray}
the family is parametrized by the $2n +2$ parameters ($b^i, \theta^i$),
which are only constrained by the condition
$\sum\limits_{i=0}^nb^i\hat{\varepsilon'}^i=0$ (so that
$\sum\limits_{i=0}^nk^i=\sum\limits_{i=0}^n\hat k^i=0$).
As a consequence
\begin{equation}
k^i\cdot k^j=-\omega^2b^ib^jcos(\theta^i-\theta^j), \qquad\qquad
\varepsilon^i\cdot k^j=-\omega b^jsin(\theta^i-\theta^j) \qquad
\varepsilon^i\cdot\varepsilon^j=-cos(\theta^i-\theta^j)
\end{equation}
\begin{equation}
\left(\varepsilon^i\cdot\varepsilon^j~k^i\cdot k^j- \varepsilon^i\cdot
 k^j~\varepsilon^j\cdot  k^i     \right)=\omega^2b^ib^j
\end{equation}

By plugging these $(k^i, \varepsilon^i)$ into Eq. (\ref{wain})  we find
\begin{eqnarray}
& & o(\omega^d)=\omega^{n+2}\sum\limits_{P'}\sum\limits_{l=1}^m
\sum\limits_{j_0,...j_{2l-1}=0}^n
\sum\limits_{i=1}^n
 \left[A^{l;~j_0...j_{2l-1}}_{i i_1...i_n}+A^{l;~j_0...j_{2l-1}}_{i_1 i...i_n}+
..+A^{l;~j_0...j_{2l-1}}_{i_1...i_{2l-1}ii_{2l}...i_n}\right]\nonumber \\
& & b^ib^{j_0}...b^{j_{2l-1}}b^{i_{2l}}...b^{i_n}cos(\theta^i-\theta^{j_0})
sin(\theta^{i_1}-\theta^{j_1})...sin(\theta^{i_{2l-1}}-\theta^{j_{2l-1}}).
\end{eqnarray}
The coefficients in the square brackets  can depend on the angles $\theta^i$
only through the cosines
$cos(\theta^i-\theta^j)$ (since
$k^i\cdot k^j=-\omega^2b^ib^jcos(\theta^i-\theta^j)$); since the above
equation has to hold for all $\theta^i$'s then all terms in the square brackets
have to satisfy the relation
\begin{equation}
 \left[A^{l;~j_0...j_{2l-1}}_{i i_1...i_n}+A^{l;~j_0...j_{2l-1}}_{i_1
i...i_n}+...+
A^{l;~j_0...j_{2l-1}}
_{i_1...i_{2l-1}ii_{2l}...i_n}\right]=o(\omega^{d-n-2})\qquad\qquad l=1,...,m
\end{equation}
independently.

Replacing the above results in formula (\ref{expr}) we find the factorization
formula
\begin{equation}
\Gamma_{n+1}^{\varepsilon^0...\varepsilon^n}(k^0,...,k^n)
 =\sum\limits_P
A^{0;}_{i_0i_1...i_n}E^{i_0i_1}...E^{i_{n-1}i_n}+o(\omega^{d-1})
\label{fact}
\end{equation}
whence formulae (\ref{fact1}), (\ref{fact2}), (\ref{fact3}) follow.

\vskip1truecm

{\bf Part 3:} On any configuration $(k^i(\omega),
\varepsilon^i(\omega))_{i=0,...,n}$
$\omega$-converging
to the decay configuration $( \hat{k}^i, \hat{\varepsilon}^i)_{i=0,...,n}$ we
have
 $E^{ij}=o(\omega^2)$.
To prove formulae (\ref{main1}), (\ref{main2}), (\ref{main3s})  it remains to
show that
the scalar functions $A$'s appearing in eq.'s (\ref{fact1}), (\ref{fact2}),
(\ref{fact3})
can show
poles in $\omega$ at most of degree  so high to yield the global
$\omega$-dependence reported in the
former formulae. For this purpose we use a continuity argument,
i.e. we argue that the claimed $\omega$-dependence is the only  one
compatible with equations
(\ref{fact1}), (\ref{fact2}), (\ref{fact3}) if we require the LHS to be
independent
of the particular configuration $(k^i(\omega),
\varepsilon^i(\omega))_{i=0,...,n}$
$\omega$-converging
to the decay configuration $( \hat{k}^i, \hat{\varepsilon}^i)_{i=0,...,n}$.

For the sake of brevity we continue to use the factorization formula
(\ref{fact}) to
deal at once
with all three cases. We choose two different multi-parameter families
 $(k^i(\omega),\varepsilon^i(\omega))$,
$(\tilde k^i(\omega),\tilde{\varepsilon}^i(\omega))$ of configurations
$\omega$-converging to the decay configuration, and we require that
\begin{equation}
\lim_{\omega\rightarrow 0}\Gamma_{n+1}^{e^0(k^0)...e^n(k^n)}(^0,...,k^n)=
\lim_{\omega\rightarrow 0}\Gamma_{n+1}^{\tilde e^0(k^0)...\tilde e^n(q^n)}
(\tilde k^0,...,\tilde k^n)
\label{ineq}
\end{equation}
In the $xyz$ axes as before, the first is the family (\ref{fam1}), the second
is
\begin{equation}
\tilde k^i:= \hat{k}^i+\omega (0,0,c^i,0)\qquad\qquad
\tilde \varepsilon^i:=[(\hat k^i_3)^2+(\omega c^i)^2]^{-\frac 12}(0,0,\hat
k^i_3,
\omega c^i)
\qquad\qquad i=0,1,...,n,
\label{fam2}
\end{equation}
where $\sum\limits^n_{i=0}c^i=0$. This implies in particular
$\tilde k^i\cdot \tilde k^j=-\omega^2c^ic^j$.

With the first family we find
\begin{equation}
\Gamma_{n+1}^{\varepsilon^0...\varepsilon^n}(k^0,...,k^n)
 =\omega^{n+1}b^0...b^n\sum\limits_P A^{0;}_{i_0i_1...i_n}+o(\omega^{d-1}).
\end{equation}

Now we specialize our discussion to the case of QED and QG, where
$d-1\ge 1$, so that the
second term vanishes when $\omega\rightarrow 0$.
Let us consider {\it per absurdum}
the hypothesis that  the  functions $A$'s have poles of degree $(n\!+\!1)$ in
$\omega$.
In order that the RHS has a limit independent of the $b^i$'s when
$\omega\rightarrow 0$,
the $A$'s must have the form
\begin{equation}
A^{0;}_{i_0i_1...i_n}=\left[ \sum\limits_P a_{i_0i_1...i_n}k^{i_0}\cdot
k^{i_1}...
k^{i_{n-1}}\cdot k^{i_n}\right]^{-1},
\label{form}
\end{equation}
where $a_{i_0i_1...i_n}$ are constants, so that
\begin{equation}
A^{0;}_{i_0i_1...i_n}=\left[\omega^{n+1}b_0...b^n\right]^{-1}\times const.
\end{equation}

On the other hand, plugging the family (\ref{fam2}) into eq. (\ref{form}) and
replacing the result into
formula (\ref{fact}), we find
\begin{equation}
\Gamma_{n+1}^{\tilde e^0(k^0)...\tilde e^n(q^n)}(\tilde k^0,...,\tilde k^n)=
const.\times
\sum\limits_P\left(\frac{d^{i_0}}{d^{i_1}}+\frac{d^{i_1}}{d^{i_0}}-1\right)...
\left(\frac{d^{i_{n-1}}}{d^{i_n}}+\frac{d^{i_n}}{d^{i_{n-1}}}-1\right)+
o(\omega^2)
\end{equation}
where we have defined $d^i:=\frac {k^i_3}{c^i}$.  This expression depends on
the choice of the
coefficients $c^i$, i.e. depends on the way the family $(\tilde k^i(\omega),
\tilde{\varepsilon}^i(\omega))$
approaches $(\hat k^i(\omega),\hat{\varepsilon}^i(\omega))$, against the
hypothesis. In a similar way,
one can exclude the hypothesis that the functions $A$'s have poles in $\omega$
of degree  $>(n\!+\!1)$, otherwise the RHS would diverge to either $+ \infty$
or
$- \infty$ according to the way the families approach the decay configuration

Summing up, we have discarded the possibility that the $A$'s have poles in
$\omega$ of degree
$\ge n\!+\! 1$, so that consequently in QED,QG
\begin{equation}
\Gamma_{n+1}^{\varepsilon^0...\varepsilon^n}(k^0,...,k^n)
 =o(\omega)
\label{bound}
\end{equation}

In QED we can improve the bound (\ref{bound}) into the  stronger bound
(\ref{main1}). In fact, if one
plugs the general expansion (\ref{stru2}) into eq. (\ref{waiqedbis}) [instead
of
eq.
(\ref{wai?})] and
argues as in part 2, one ends up with a stronger form of the factorization,
\begin{equation}
\Gamma_{n+1}^{\mu_0....\mu_n}(k^0,...,k^n)
=\sum\limits_P
A_{i_0i_1...i_n}E^{\mu_{i_0}\mu_{i_1}}...E^{\mu_{i_{n-1}}\mu_{i_n}}.
\label{factbis}
\end{equation}
Looking at the Feynman diagrams contributing to each order in the loops to
$\Gamma_{n+1}^{\mu_0...\mu_n}(k^0,...,k^n)$, it is easily realized that 
they are continuous and finite for all values of $k^i$'s, since the 
fermion/scalar masses are infrared cutoffs [see fig.\ (\ref{box})].
Hence, the scalars $A$ cannot have poles in $k^i\cdot k^j$, because 
otherwise at least the terms $A_{i_0i_1...i_n}(k^{i_0})^{\mu_{i_1}}...
(k^{i_n})^{\mu_{i_{n-1}}}(k^{i_{n-1}})^{\mu_{i_n}}$ would
diverge. The $A$'s have dimension $[mass]^{4-2(n+1)}$, since
$\Gamma_{n+1}$ has dimension $[mass]^{4-(n+1)}$. This can be accounted 
for without introducing poles in
$k^i\cdot k^j$, but using the mass parameters of the charged particle
interacting with the photon. For instance, if the only charged particle is a
fermion with mass $m$, then $A= m^{4-2(n+1)}o(1)$. We have completed
the proof of the claim (\ref{main1}).

In QG the $o(\omega)$ in the RHS of  (\ref{bound}) can be improved into a
$o(\omega^2)$, since
$\Gamma_{n+1}^{e^0...e^n}(k^0,...,k^n)$ can be only of even degree in
$\omega$, if we
assume
that the proper vertices depend analitically on the momenta $k^i$. This follows
from formula
(\ref{approx}),  because the LHS of eq.  (\ref{bound})  has
to be a function of the Lorentz scalars $k^i\cdot k^j$,
$\varepsilon^i(k^i)\cdot k^j$, of even degree
in the latter. This completes the proof of the claim (\ref{main3s}).

In YM formula (\ref{fact}) and the continuity argument do not exclude that
there exists a limit $\lim\limits_{\omega \rightarrow 0}
\Gamma_{n+1}^{\varepsilon^0...\varepsilon^n}(k^0,...,k^n)=:L\neq 0$
independent of the way the family $(k^i(\omega),\varepsilon^i(\omega))$
approaches $(\hat k^i(\omega),\hat{\varepsilon}^i(\omega))$.
 In fact, if the functions $A$'s have a pole of degree $\ge (n\!+\!1)$ in
$\omega$,
the second term in formula (\ref{fact}) (which in principle can be finite
or divergent) could  compete with
the first, and $\Gamma_{n+1}^{\varepsilon^0...\varepsilon^n}(k^0,...,k^n)$
could
have
a family-independent limit even though the first term has not. This is exactly
what happens
with the 4-gluon proper vertex, as one can already check at the tree level
\begin{equation}
\Gamma_{4,tree}^{\hat{\varepsilon}^0a_0...\hat{\varepsilon}^3a_3}\propto
\left[ (\hat{ \varepsilon}^0\cdot \hat{\varepsilon}^1)
  (\hat{\varepsilon}^2\cdot \hat{\varepsilon}^3) f^{a_0a_3e}f^{a_1a_2e}+ perm.
\right]\neq 0.
\end{equation}
By an explicit analysis of the general expansion (\ref{stru2}) one can easily
realize that a
family-independent limit $L\in {\bf R}\cup\{\pm \infty\}$ can be obtained only
if
equation
(\ref{main2}) is satisfied.

Finally, the proof of the general claim (\ref{main3}) can be done by an
induction
procedure in the number of external photons (resp. of YM bosons) which mimics
the one sketched
so far for QED (resp. YM), with the only difference that as starting input we
do
not use the value
of proper vertex with zero photons, zero YM bosons and zero gravitons,  but
the proper
vertex with $m_g>0$ gravitons or $m_y>0$ YM bosons (resp. with $m_g>0$
gravitons
or $m_{\gamma}>0$ photons).

We have thus  completed the proof  of property 10 $\diamondsuit$.

\section{Concluding remarks.}

 We have seen that the decay probabilities for the photon, the
 graviton and the Yang-Mills boson all vanish (perturbatively).
 The decay amplitudes
 involving only photons and/or gravitons are themselves zero;
 we have first shown these properties by a simple power counting
 argument and then proved them rigorously through the Ward identities,
 assuming only continuity of the Greens functions in the infrared
 limit. In the case of the Yang-Mills boson,
 the amplitude does not vanish in the infrared limit (more precisely,
 it diverges if $m\ge 5$ out of $n\!+\!1$ external
 particles are YM bosons);
 the decay probability is however suppressed by the phase-space factor.
 The latter is the only case in which we have needed an infrared regulator.
 
 In this final Section we would like to comment on the relation between our 
 work and the classical literature \cite{ir}
 on infrared divergences in quantum field theory.
 
 For the reasons just mentioned, even in YM theories we do not need to 
 average ({\it \`a la  Bloch and Nordsieck} \cite{ir}) over sets of states
 degenerate in the energy, like in the Kinoshita-Lee-Nauenberg theorem 
 \cite{ir}, in order to build finite physical transition probabilities 
 out of divergent amplitudes.
 
 However, this
 might be necessary for other theories, not explicitly considered
 here, where the divergences of the amplitudes are sufficiently bad.
 In the latter case it would be nevertheless important to keep in mind
 some peculiarities of the decay of massless particles compared to what
 one usually finds in the literature \cite{ir}.
 The physical processes explicitly considered in the literature
 are either scatterings, or decays in which the initial particle is 
 massive. The case of a decay process where all particles (including 
 the initial one) are massless is not considered. In a theory 
 including massless particles, the
 Green functions may diverge if (1) some of the external particles are 
 soft/collinear massless ones and/or (2) if massless particles appear among 
 the internal ones occurring in the corresponding Feynman diagrams
 (e.g. in loops).
 The study of these divergences is usually performed by attributing a small
 mass $m$ to each kind of massless particle in the theory and then studying 
 the limit in which $m$ goes to zero (as already recalled, their
 elimination from
 "physical" transition probabilities is obtained by building up the initial
 and/or final states as a mixture of degenerate states of the energy 
 before performing the limit; see Ref.'s \cite{ir}).
 
 On the other hand we note that, while in the scattering processes or in
 the decay processes of a massive particle the collinearity of some 
 massless external particles is one of the kinematically allowed
 configuration [so that it makes sense to study the divergences of the
 Green functions in the limit
 when these external momenta become collinear while remaining on-shell
 (null)], in the decay of a massless particle the only
 allowed kinematical configuration is that in which all the external 
 particles are collinear. Therefore, the divergences, if they occur,
 characterize all the kinematically allowed configurations. Moreover,
 the latter tipically appear already at the tree-level (consider
 e.g. $T_5$ for the YM theory), as one can immediately
 check by summing all relevant tree diagrams.
 As for the IR regulator, the one based on the attribution of a small
 mass to the external particles is unsuitable because
 it forbids the decay of a particle into other ones of the
 same kind (as we have already noted in section \ref{regu}).
 Our regulator, based on a small-frequency external source, bypasses
 this difficulty while having the nice feature of being physically
 intuitive.
 
 \medskip
 As mentioned at the end of Section 2,
 a partial decay probability $\Gamma_n$ different from zero can
 be only obtained when the square amplitude is proportional to a
 sufficiently high negative power of $\omega$. If we admit (as
 is generally true in perturbation theory) that the coupling constants
 appear in the numerator, this means that the amplitude must contain
 a coupling constant with positive mass dimension.
 
 One of the few theories we are aware of, in which such a coupling occurs
 (besides the $\lambda \phi^3$ theory; compare Section 1)
 is gravity in the presence of a cosmological constant. In this case
 the action of the gravitational field is written as
 \begin{equation}
   S = \frac{1}{16 \pi G} \int d^4x \, \sqrt{g(x)} \, [\Lambda - R(x)]
 \end{equation}
 or, redefining the metric in the form $g_{\mu \nu}(x) = \eta_{\mu \nu}
 + \kappa \tilde{h}_{\mu \nu}(x)$, with $\kappa = \sqrt{16 \pi G}$,
 \begin{equation}
   S = \int d^4x \, \sqrt{1+\kappa \tilde{h} + \kappa^2 \tilde{h}^2
   + \kappa^3 \tilde{h}^3 + ... } \, \left[ \frac{\Lambda}{\kappa^2} -
   \tilde{R}^{(2)}(x) + ... \right]
 \label{pizza}
 \end{equation}
 We have denoted symbolically with $\tilde{h}, \ \tilde{h}^2, \
 \tilde{h}^3 \ ...$ in the square root terms which are linear, quadratic,
 cubic ... in $\tilde{h}$, omitting the indices and the exact algebraic
 structure. $\tilde{R}^{(2)}(x)$ denotes the part of the curvature
 quadratic in $\tilde{h}$. The term $\kappa^3 \tilde{h}^3$, when is
 multiplied by $\Lambda / \kappa^2$, gives rise to a vertex $\kappa
 \Lambda \tilde{h}^3$ which couples three gravitons with a coupling
 constant $\kappa \Lambda$ of mass
 dimension 1 (unlike the corresponding three-vertex of the pure Einstein
 action, which is proportional to $\kappa^3$ and contains 4 four-momenta,
 so that the infrared processes are strongly suppressed).
 
 Although the decay amplitudes involving this new three-vertex are
 suppressed at the tree-level because of helicity conservation (Property
 3), it can be used to construct gravitonic loops with $n$ external
 legs. The amplitudes will be proportional to positive powers of 
 $\kappa \Lambda$ and -- in our regularization
 scheme -- to negative powers of $\omega$. This means that $\Gamma_n$
 would be finite in the limit $\omega \to 0$, or even diverge.
 But we should not forget the terms which are linear and quadratic in
 $\tilde{h}$ in the square root of eq.\ (\ref{pizza}). In particular,
 the quadratic term gives rise to a graviton mass (if $\Lambda < 0$)
 or to instability (if $\Lambda >0$) \cite{vel}. In the
 first case, we end up with gravitons which are not massless any more,
 so that all our preceding formalism does not apply.
 
 It is known that the cosmological constant $\Lambda$,
 although possibly very large in principle, is limited by astronomical
 observations to be less than $|\Lambda| \leq 10^{120} G^{-1}$ (in order 
 to explain this vanishing, many mechanisms have been proposed \cite{wei}).
 Therefore it seems that the idea of a decay induced by the presence of a
 cosmological constant can be excluded on the basis of the empirical
 evidence.
 
 However, in the non-perturbative quantum Regge calculus \cite{ham}
 the effective value of the adimensional product $|\Lambda| G$ depends
 on the length scale and vanishes according to a power law as the 
 energy scale $\mu$ goes to zero:
 \begin{equation}
   |\Lambda|G \sim (l_0 \mu)^\gamma
 \end{equation}
 If we admit that the average lattice spacing $l_0$ is of the order of 
 the Planck length \cite{mod}, then the constant $\Lambda$ can be 
 non-vanishing on small scales, leaving the graviton massless at
 large scales. This might change the situation, but clearly
 at the present stage of knowledge these are still speculative hypotheses.
 
 \medskip
 
 We wish to thank M.\ Abud, P.\ Van Nieuwenhuizen and G.\ A.\ Vilkovisky
 for useful discussions, as well as the A. Von Humboldt foundation for financial
 support. We are grateful to D. Maison and J. Wess for the kind
 hospitality at their institutes.

\section*{Appendix}

We prove eq. (\ref{appen}). In the case of QED with (for instance)
Feynman's gauge-fixing $\frac 1{2\alpha}\int d^4x (\partial_{\mu}A_{\mu})^2$,
the LHS is zero when $n>1$ because the gauge variation of the gauge-fixing
above is of first degree in $A_{\mu}$, and is zero in the case $n=1$ because
\begin{equation}
-\frac 1{\alpha}p_1^{\mu}(p_1)^2\xi(p_1),
\label{prec}
\end{equation}
vanishes after contraction with the polarization vector 
$\varepsilon^{\pm}_{\mu}(p_1)$. In the case of YM with (for instance) 
Feynman's gauge-fixing $\frac 1{\alpha}\int d^4x (\partial^{\mu}A_{\mu})^2$,
the LHS is zero if $n>2$ because the gauge variation of the gauge-fixing above
is of second degree in $A_{\mu}$; if $n=1$ it is zero for the same reason as
in the preceding case (\ref{prec}); if $n=2$ it is zero because
\begin{equation}
-\frac 2{\alpha}(p_1)^{\mu_1}(p_2)^{\mu_2}\xi^c(p_1+p_2)f^{a_1a_2c},
\end{equation}
vanishes after contractions with the polarization vector
$\varepsilon^{\pm}_{\mu_1}(p_1)\varepsilon^{\pm}_{\mu_2}(p_2)$.
In the case of QG with harmonic gauge-fixing
$\frac 1{2\alpha}\int d^4x (\partial^{\mu}h_{\mu\nu})^2$ we have
\begin{equation}
\delta_{\xi}\left[\frac 1{2\alpha}\int d^4x (\partial^{\mu}h_{\mu\nu})^2
\right]=-\frac 1{\alpha}\int d^4p (p^{\mu}h_{\mu\nu}(p))
p^{\rho}(\hat{\xi_{;\rho}^{\nu}}+\hat{\xi^{;\nu}_{\rho}})(-p)
\end{equation}
(here $\hat{~}$ means Fourier transform). When some $\frac{\delta\qquad}
{\delta g_{\mu_i\nu_i}(-p_i)}$ acts on $p^{\mu}h_{\mu\nu}(p)$ we get a factor
$\delta^4(p+p_i)[\delta^{\mu_i}_{\mu}\delta^{\nu_i}_{\nu}+\delta^{\mu_i}_{\nu}
\delta^{\nu_i}_{\mu}]$ (see formula (\ref{deriv})),
which gives zero after contraction with the polarization tensor
$e^{\mu_i\nu_i}(p_i)$. $\diamondsuit$

\end{document}